\documentclass[aps,pre,reprint,superscriptaddress,nofootinbib]{revtex4-2}

\usepackage{amsmath}
\usepackage{amssymb}
\usepackage{bm}
\usepackage{graphicx}
\usepackage[colorlinks=true,allcolors=blue]{hyperref}
\usepackage{comment}

\begin{document}

\title{Geometry-Controlled Initial-Density Dependence in Branching-Merging Exclusion Processes}

\author{Hiroki Yamamoto}
\affiliation{Chiba Municipal Aoba Hospital, 1273-2 Aoba-cho, Chuo-ku, Chiba 260-0852, Japan}

\author{Daichi Yanagisawa}
\affiliation{Department of Aeronautics and Astronautics, School of Engineering, The University of Tokyo,\\7-3-1 Hongo, Bunkyo-ku, Tokyo 113-8656, Japan}
\affiliation{Department of Advanced Interdisciplinary Studies, School of Engineering, The University of Tokyo, \\4-6-1, Komaba, Meguro-ku, Tokyo, 153-8904, Japan}
\affiliation{Mobility Innovation Collaborative Research Organization, The University of Tokyo,\\5-1-5, Kashiwanoha, Kashiwa-shi, Chiba, 277-8574, Japan}

\author{Katsuhiro Nishinari}
\altaffiliation[Also at ]{Research Center for Advanced Science and Technology, The University of Tokyo, 4-6-1 Komaba, Meguro-ku, Tokyo 153-8904, Japan}
\affiliation{Department of Aeronautics and Astronautics, School of Engineering, The University of Tokyo,\\7-3-1 Hongo, Bunkyo-ku, Tokyo 113-8656, Japan}
\affiliation{Mobility Innovation Collaborative Research Organization, The University of Tokyo,\\5-1-5, Kashiwanoha, Kashiwa-shi, Chiba, 277-8574, Japan}
\affiliation{Department of Advanced Interdisciplinary Studies, School of Engineering, The University of Tokyo,\\4-6-1, Komaba, Meguro-ku, Tokyo, 153-8904, Japan}

\date{\today}

\begin{abstract}
We study geometry-controlled, initial-density-dependent stationary states
in exclusion processes with branching-merging structures.  We consider two
boundary geometries for connecting the parallel branches to the reservoirs:
the \(1\)-\(n\)-\(1\) model, in which a single lane branches into \(n\)
parallel lanes and then merges back into a single lane, and the
\(0\)-\(n\)-\(0\) model, in which particles directly enter and leave the
\(n\) parallel lanes.  For both models, we derive critical initial
densities separating the low-density, shock-wave, and high-density
regimes, and show that the theoretical predictions reproduce the phase
diagrams and branch-averaged densities observed in simulations.  In the
\(1\)-\(n\)-\(1\) geometry, the critical initial densities depend on both
the number of branches and the lengths of the upstream and downstream
single-lane segments, whereas in the \(0\)-\(n\)-\(0\) geometry the
corresponding estimates depend only on the number of branches.  These
results show that initial-condition dependence can arise from the global
geometry of a branching-merging exclusion process, even without an
explicit friction rule at the merging point.
\end{abstract}

\maketitle

\section{Introduction}
\label{sec:introduction}

The totally asymmetric simple exclusion process (TASEP) is a paradigmatic
model of nonequilibrium transport. Since its introduction by MacDonald
and Gibbs~\cite{macdonald1968kinetics,macdonald1969concerning}, the TASEP
and its variants have been applied to vehicular traffic, molecular
motors, and transport in biological and social systems
~\cite{chowdhury2000statistical,chowdhury2005physics,schadschneider2010stochastic}.
In the standard open-boundary TASEP, the stationary state is selected by the entrance and exit probabilities. Under fully parallel update, where all particles are updated simultaneously, for example, the system exhibits low-density (LD), high-density (HD), and shock-wave (SW) phases~\cite{de1999exact}. For fixed boundary probabilities, the phase is usually independent of the initial condition.

Transport on networks provides an important extension of the TASEP. Network versions of exclusion processes have been used to describe
transport on complex graphs and cytoskeletal networks~\cite{neri2011totally}, as well as systems with branching, merging, and junction structures
~\cite{PhysRevE.69.066128,PhysRevE.77.051108,LIU20094068}. Such geometries can create bottlenecks and strongly modify stationary flows and density profiles. Related forked and double-chain lattices also exhibit nontrivial density profiles and flow-optimization effects~\cite{PhysRevE.87.062116,PhysRevE.91.062818}. In most of these systems, however, the long-time state is determined by boundary probabilities and network geometry, rather than by the initial density.

Initial-condition dependence is closely related to metastability and effective ergodicity breaking. In traffic-flow models and exclusion processes, metastable states have been reported in systems with
slow-to-start rules or related dynamical mechanisms
~\cite{PhysRevE.55.5597,barlovic1998metastable,PhysRevLett.86.2498,PhysRevE.66.046113,Nishinari_2004}.
Ergodicity breaking has also been discussed in exclusion processes with condensation or varying system length~\cite{PhysRevLett.89.090601,SCHULTENS2015100}.
These studies show that memory of the initial state can persist when the dynamics stabilizes distinct macroscopic states.

Recently, metastability and initial-condition dependence have been reported in branching-merging exclusion processes when additional mechanisms, such as junction friction or resource constraints, are introduced~\cite{PhysRevE.108.044121,bhatia2026emerge}.  These studies suggest that conflicts at network junctions can stabilize distinct stationary
states.

In this paper, we address this question by studying two minimal multilane
TASEP geometries. The first is the \(1\)-\(n\)-\(1\) model, in which a
single upstream lane branches into \(n\) parallel lanes and then merges
back into a single downstream lane. The second is the \(0\)-\(n\)-\(0\)
model, in which the \(n\) parallel lanes are directly connected to the entrance and exit boundaries.

We show that both models exhibit initial-density-dependent macroscopic
states: depending on the initial density, the system relaxes to the LD
phase, the HD phase, or a persistent local SW phase.  The critical
initial densities are controlled by geometry in two senses.  In the
$1$-$n$-$1$ model, they vary with the lengths of the upstream and
downstream single-lane segments as well as with the number of branches.
Moreover, comparing the $1$-$n$-$1$ and $0$-$n$-$0$ models shows that the
way in which the branches are connected to the boundaries changes the
critical densities and density profiles.  Thus, the persistence and
spatial manifestation of initial-condition dependence are controlled not
only by the initial density itself, but also by the global boundary
geometry.

The remainder of this paper is organized as follows.
Section~\ref{sec:model} defines the \(1\)-\(n\)-\(1\) and \(0\)-\(n\)-\(0\)
models. Section~\ref{sec:analyses1n1} analyzes the $1$-$n$-$1$ model, focusing
on how the initial density and the lane lengths determine the phase
diagram, mean branch density, density profiles, and local shock-wave picture. Section~\ref{sec:analyses0n0}
then applies the same analysis to the \(0\)-\(n\)-\(0\) model. Finally, Sec.~\ref{sec:conclusion} summarizes the results and discusses their
implications.

\section{Model}
\label{sec:model}

\begin{figure}[t]
\centering
\includegraphics[width=\columnwidth]{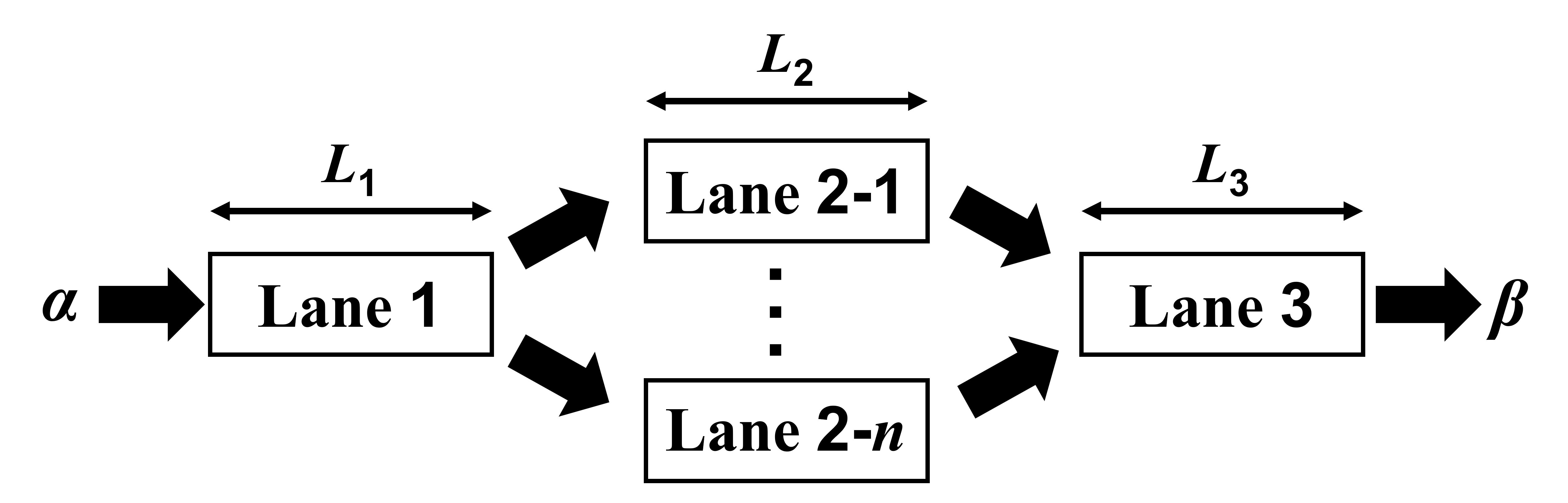}
\caption{
Schematic illustration of the $1$-$n$-$1$ model. The system consists of one upstream lane of length $L_1$, $n$ branches of length $L_2$, and one downstream lane of length $L_3$.
}
\label{fig:1n1model}
\end{figure}

\begin{figure}[t]
\centering
\includegraphics[width=\columnwidth]{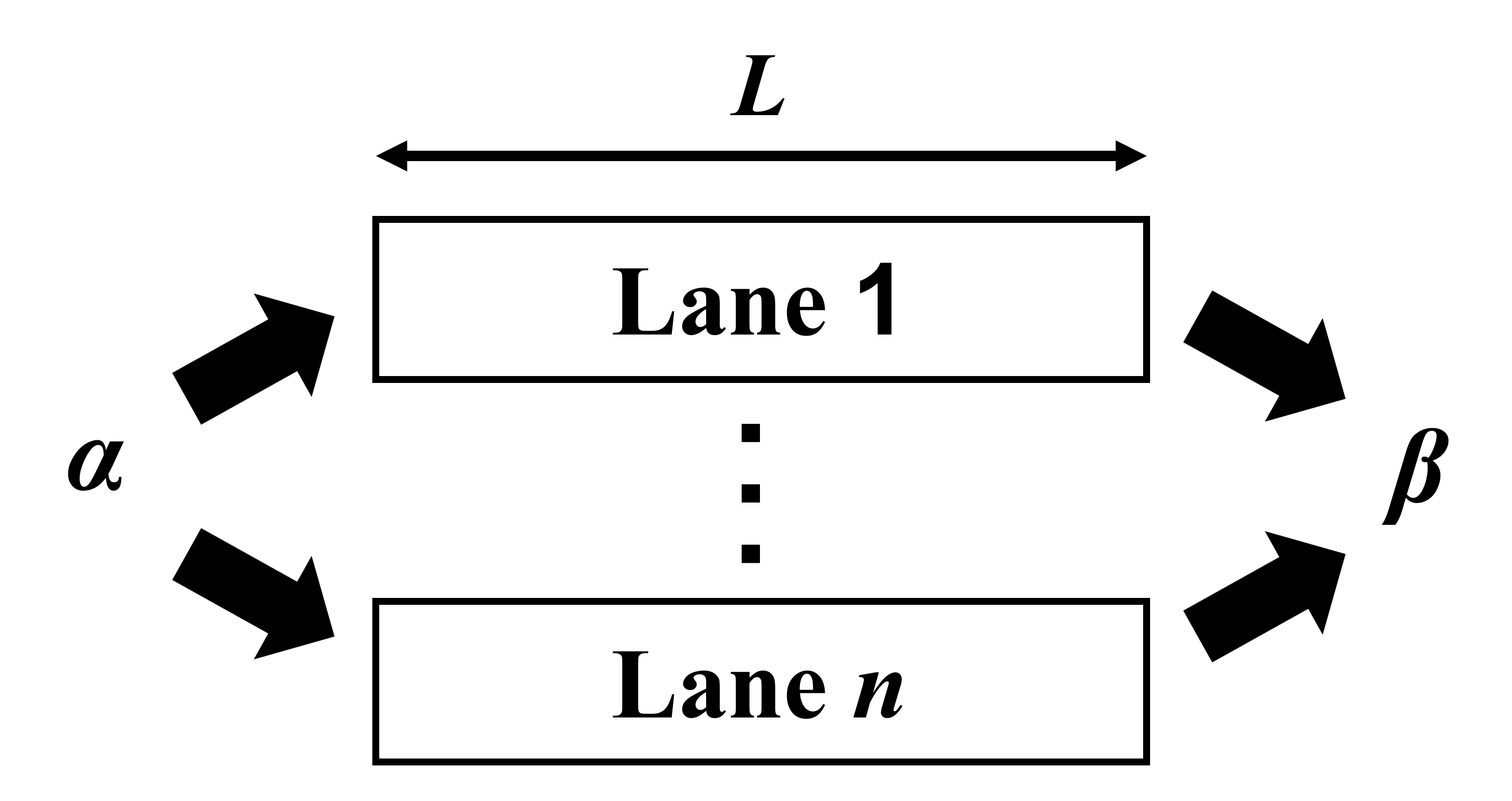}
\caption{
Schematic illustration of the $0$-$n$-$0$ model. The system consists only of $n$ parallel branches of length $L$.
}
\label{fig:0n0model}
\end{figure}

We consider two related multilane TASEP models: the $1$-$n$-$1$ model and the $0$-$n$-$0$ model, as shown in Figs.~\ref{fig:1n1model} and \ref{fig:0n0model}, respectively. In both models, particles obey the exclusion rule; each site can be either empty or occupied by at most one particle. Parallel updating with discrete time is employed, signifying that all particles are updated simultaneously. The bulk hopping probability is set to 1 throughout this study. Thus, a particle moves to the right-neighboring site if that site is empty; otherwise, it remains at its present site.

\subsection{$1$-$n$-$1$ model}

The $1$-$n$-$1$ model consists of one upstream lane (Lane 1), $n$ parallel branches (Lanes 2-1, $\ldots$, 2-$n$), and one downstream lane (Lane 3). The lengths of Lane 1, each branch in Lanes 2, and Lane 3 are denoted by $L_1$, $L_2$, and $L_3$, respectively. 

Initially, particles are randomly placed with density $\rho_{\rm ini}$. A particle is injected into Lane 1 from the left boundary with probability $\alpha$ when the first site of Lane 1 is empty. A particle at the last site of Lane 1 attempts to enter Lanes 2. If at least one first site of Lanes 2 is empty, the particle moves into one of the available branches; when multiple branches are available, one of them is selected randomly.
At the merging point, particles at the last sites of the branches can move into Lane 3. If the first site of Lane 3 is empty and more than one branch has a particle at its last site, one of these particles is selected randomly and moves into Lane 3. A particle at the last site of Lane 3 leaves the system with probability $\beta$.

\subsection{$0$-$n$-$0$ model}

The $0$-$n$-$0$ model consists only of $n$ parallel branches (Lanes 1, $\ldots$, $n$). There is no upstream lane and no downstream lane after them. The length of each branch is denoted by $L$.

The particles were placed uniformly over all branch sites with $\rho_{\rm ini}$. A particle is directly injected into the branches from the left boundary with probability $\alpha$. If at least one first site of the branches is empty, the particle enters one of the available branches; when multiple branches are available, one of them is selected randomly. At the right boundary, if at least one last site of the branches is occupied, one particle leaves the system with probability $\beta$. When multiple particles to leave are available at the last sites, one of them is selected randomly.

\par\vspace{1em}

For both models, we define \(x\in[0,1]\) as the normalized position along a branch, measured from the upstream end (\(x=0\)) to the downstream end (\(x=1\)). In the main text, we set $\alpha=\beta=1$.  In the simulations, the system is
first relaxed for $10^6$ time steps, and physical quantities are then obtained by averaging over $10^6$ time steps, otherwise specified. In addition, we obtain the simulation results by averaging ten trials, also otherwise specified.

\section{Analyses of $1$-$n$-$1$ model}
\label{sec:analyses1n1}

\subsection{Phase Diagram and Mean Branch Density}
\label{sec:PD1n1}

We first analyze how the initial density and the lane lengths determine the stationary state of the branches in the $1$-$n$-$1$ model. The key question is whether the particles initially existing in the relevant region can leave before newly injected particles catch up with them. If they can leave, the system relaxes to the LD state. If some of them remain, the incoming particles interact with the remaining high-density region, and a local LD--HD coexistence layer is formed.


The lower transition is determined by the early-time
catching-up process of particles injected from the upstream side. In this process, the value $L_3$ does not influence the leading estimate.

The first particle injected from the left boundary reaches the end of
the branching region after a time of order
\begin{equation}
t_{\rm arr}
\simeq
L_1+L_2 .
\label{eq:t_arr_1n1}
\end{equation}
The number of particles initially existing in the upstream lane and the
$n$ branches is
\begin{equation}
N_{\rm ini}
=
\rho_{\rm ini}
\left(
L_1+nL_2
\right).
\label{eq:N_ini_1n1}
\end{equation}

We next estimate the number of particles that can leave the
upstream-and-branching region before the first newly injected particle
reaches the end of the branching region. For low initial density, the
total exit flow from the branches is proportional to the number of
branches and to $\rho_{\rm ini}$. However, because the particles leaving
the branches must enter the single downstream lane, the total flow is
bounded by 1/2, which corresponds to the maximal flow of the downstream lane. Thus, the
early-time exit flow is estimated as
\begin{equation}
Q_{\rm ini,ex}
=
\begin{cases}
n\rho_{\rm ini}
& {\rm for \ } 0\leq\rho_{\rm ini}<\dfrac{1}{2n},\\
\dfrac{1}{2}
& {\rm for \ }  \dfrac{1}{2n}\leq\rho_{\rm ini}\leq\dfrac{1}{2}.
\end{cases}
\label{eq:Q_ini_ex_1n1}
\end{equation}
Therefore, the number of particles that can leave the
upstream-and-branching region during the time interval $t_{\rm arr}$ is
\begin{equation}
N_{\rm ex}
\simeq
Q_{\rm ini,ex}t_{\rm arr},
\label{eq:N_ex_1n1_def}
\end{equation}
or explicitly,
\begin{equation}
N_{\rm ex}
=
\begin{cases}
n\rho_{\rm ini}
\left(
L_1+L_2
\right)
&{\rm for \ }  0\leq\rho_{\rm ini}<\dfrac{1}{2n},\\
\dfrac{1}{2}
\left(
L_1+L_2
\right)
&{\rm for \ }  \dfrac{1}{2n}\leq\rho_{\rm ini}\leq\dfrac{1}{2}.
\end{cases}
\label{eq:N_ex_1n1}
\end{equation}

The lower critical condition is obtained by comparing $N_{\rm ini}$ and
$N_{\rm ex}$. For $0\leq\rho_{\rm ini}<1/(2n)$, we have
\begin{align}
N_{\rm ex}-N_{\rm ini}
&=
n\rho_{\rm ini}
\left(
L_1+L_2
\right)
-
\rho_{\rm ini}
\left(
L_1+nL_2
\right)
\nonumber\\
&=
\rho_{\rm ini}
\left(
n-1
\right)
L_1 .
\label{eq:clearance_1n1}
\end{align}
This quantity is positive for $n>1$. Therefore, in this region, the initially existing particles are removed from the system before the newly injected particles catch up, with a finite margin.

For $1/(2n)\leq\rho_{\rm ini}\leq1/2$, the exit flow is saturated. In
this region, the condition for local coexistence-layer formation is
\begin{equation}
N_{\rm ini}>N_{\rm ex}.
\label{eq:SW_condition_1n1}
\end{equation}
Using Eqs.~\eqref{eq:N_ini_1n1} and \eqref{eq:N_ex_1n1}, this condition
becomes
\begin{equation}
\rho_{\rm ini}
\left(
L_1+nL_2
\right)
>
\frac{1}{2}
\left(
L_1+L_2
\right).
\label{eq:SW_condition_1n1_expanded}
\end{equation}
Thus, the lower critical initial density is
\begin{equation}
\rho_-
=
\frac{L_1+L_2}
{2\left(L_1+nL_2\right)}
=
\frac{r_1+1}
{2\left(r_1+n\right)},
\label{eq:rho_minus_1n1}
\end{equation}
where $r_1=L_1/L_2$. The branches are expected to exhibit the LD state
for $\rho_{\rm ini}<\rho_-$, whereas a local coexistence layer appears
for $\rho_{\rm ini}>\rho_-$.

The high-density boundary is obtained by applying the same argument to
holes. Under particle-hole transformation with spatial reversal, holes
move from the downstream side toward the upstream side. Therefore, the
role of the upstream length for holes is played by $L_3$, not by $L_1$.
The first hole injected from the right boundary reaches the beginning of
the branching region after a time of order
\begin{equation}
t_{\rm arr}^{\rm h}
\simeq
L_3+L_2 .
\label{eq:t_arr_hole_1n1}
\end{equation}
If $\rho_{\rm h}=1-\rho_{\rm ini}$ denotes the initial hole density, the
same estimate gives the lower critical density for holes as
\begin{equation}
\rho_{{\rm h},-}
=
\frac{L_3+L_2}
{2\left(L_3+nL_2\right)}
=
\frac{r_3+1}
{2\left(r_3+n\right)},
\label{eq:rho_hole_minus_1n1}
\end{equation}
where $r_3=L_3/L_2$. Transforming back to the particle density, the
upper critical density is
\begin{equation}
\rho_+
=
1-\rho_{{\rm h},-}
=
1-
\frac{r_3+1}
{2\left(r_3+n\right)} .
\label{eq:rho_plus_1n1}
\end{equation}

Thus, the lower boundary is controlled by $L_1$,
whereas the upper boundary is controlled by $L_3$.
The simple relation $\rho_+=1-\rho_-$ holds only for the symmetric case of $L_1=L_3$. 

The expected phases are summarized as
\begin{equation}
\begin{cases}
{\rm LD}
& {\rm for \ } 0\leq\rho_{\rm ini}<\rho_-,\\
{\rm SW}
& {\rm for \ } \rho_-\leq\rho_{\rm ini}\leq\rho_+,\\
{\rm HD}
& {\rm for \ } \rho_+<\rho_{\rm ini}\leq1.
\end{cases}
\label{eq:phase_1n1}
\end{equation}
Here, the SW phase denotes a state in which a shock wave, or more
precisely a local LD--HD coexistence layer, is formed in the branching
region. As described below, this coexistence
layer does not generally wander over the whole system in contrast to the usual domain-wall picture. 

We next estimate the branch-averaged density $\bar{\rho}_{\rm br}$ from particle-number balance. The values of $\bar{\rho}_{\rm br}$ in the LD and HD phases in each
branch are first considered.  In the LD and HD phases, the branch density is spatially uniform in the bulk and is determined by the branching flow in the LD phase and by the merging flow in the HD phase. The total current through the downstream
single lane is bounded by the maximal-flow value $1/2$.
This flow is shared by the $n$ branches; hence $\bar{\rho}_{\rm br}$ are estimated as
\begin{equation}
\bar{\rho}_{\rm br}=
\begin{cases}
\dfrac{1}{2n}(=\rho_{\rm L})
& {\rm for \ LD},\\
1-\dfrac{1}{2n}(=\rho_{\rm H})
& {\rm for \ HD}.
\end{cases}
\label{eq:rho_LH_1n1}
\end{equation}

Then, we consider $\bar{\rho}_{\rm br}$ in the SW phase on the lower-density side, i.e., $\rho_-\leq\rho_{\rm ini}<1/2$, by focusing on the particle balance in the branching region. Since the catching-up time is given by Eq.~\eqref{eq:t_arr_1n1}, about $t_{\rm arr}/2$ particles
enter the branches, which is equal to $N_{\rm ex}$.  These
two flows cancel in the estimate of the branch particle number.  The
remaining correction is that the upstream lane is effectively filled with density $1/2$ at the time of catching up.  Hence, the number of particles
remaining in the branches is estimated as
\begin{equation}
N_{\rm br}^{(-)}
=
N_{\rm ini}-\frac{L_1}{2}
=
\rho_{\rm ini}
\left(
L_1+nL_2
\right)
-
\frac{L_1}{2},
\label{eq:N_br_1n1}
\end{equation}
using Eq. (\ref{eq:N_ini_1n1}), and therefore, the corresponding branch-averaged density is
\begin{align}
\bar{\rho}_{\rm br}^{(-)}
&=
\frac{
\rho_{\rm ini}
\left(
L_1+nL_2
\right)
-
L_1/2
}
{nL_2}\nonumber\\
&=
\frac{
\rho_{\rm ini}
\left(
r_1+n
\right)
-
r_1/2
}
{n},
\label{eq:rho_bar_branch_1n1_lower}
\end{align}
which satisfies $\bar{\rho}_{\rm br}^{(-)}>0$ due to $\rho_{\rm L}>\rho_-$. 

On the higher-density side, i.e., $1/2<\rho_{\rm ini}\leq\rho_+$, the same balance is more naturally written for holes. Let $\rho_{\rm h}=1-\rho_{\rm ini}$. Lane 3 contributes $L_3/2$ holes in the corresponding maximal-current
state, and therefore the branch-averaged hole density is estimated as
\begin{align}
\bar{\rho}_{{\rm h},{\rm br}}^{(+)}
&=
\frac{
\rho_{\rm h}
\left(
L_3+nL_2
\right)
-
L_3/2
}
{nL_2}\nonumber\\
&=
\frac{
(1-\rho_{\rm ini})
\left(
r_3+n
\right)
-
r_3/2
}
{n}.
\label{eq:rho_bar_hole_branch_1n1}
\end{align}
The particle density in the branches is then
\begin{equation}
\bar{\rho}_{\rm br}^{(+)}
=
1-\bar{\rho}_{{\rm h},{\rm br}}^{(+)}.
\label{eq:rho_bar_branch_1n1_upper}
\end{equation}
Equations~\eqref{eq:rho_bar_branch_1n1_lower} and
\eqref{eq:rho_bar_branch_1n1_upper} meet at $\rho_{\rm ini}=1/2$:
\begin{equation}
\bar{\rho}_{\rm br}^{(-)}
=
\bar{\rho}_{\rm br}^{(+)}
=
\frac{1}{2}.
\label{eq:rho_half_matching_1n1}
\end{equation}
Therefore, for $L_1\neq L_3$, the two sides have
different slopes,
\begin{equation}
\frac{d\bar{\rho}_{\rm br}^{(-)}}{d\rho_{\rm ini}}
=
\frac{L_1+nL_2}{nL_2},
\qquad
\frac{d\bar{\rho}_{\rm br}^{(+)}}{d\rho_{\rm ini}}
=
\frac{L_3+nL_2}{nL_2},
\label{eq:rho_bar_slopes_1n1}
\end{equation}
but both branches of the estimate pass through
$(\rho_{\rm ini},\bar{\rho}_{\rm br})=(1/2,1/2)$. 

Eventually, $\bar{\rho}_{\rm br}$ is summarized as
\begin{widetext}
\begin{equation}
\bar{\rho}_{\rm br}=
\begin{cases}
\rho_{\rm L}
&{\rm for \ } 0\leq\rho_{\rm ini}<\rho_-,\\
\dfrac{
\rho_{\rm ini}
\left(
r_1+n
\right)
-
r_1/2
}
{n}
&{\rm for \ } \rho_-\leq\rho_{\rm ini}<\dfrac{1}{2},\\
1-\dfrac{
(1-\rho_{\rm ini})
\left(
r_3+n
\right)
-
r_3/2
}
{n}
&{\rm for \ } \dfrac{1}{2}\leq\rho_{\rm ini}<\rho_+,\\
\rho_{\rm H}
&{\rm for \ } \rho_+<\rho_{\rm ini}\leq1.
\end{cases}
\label{eq:barrhobr_1n1}
\end{equation}
\end{widetext}

\begin{figure*}[t]
\centering
\includegraphics[width=\textwidth]{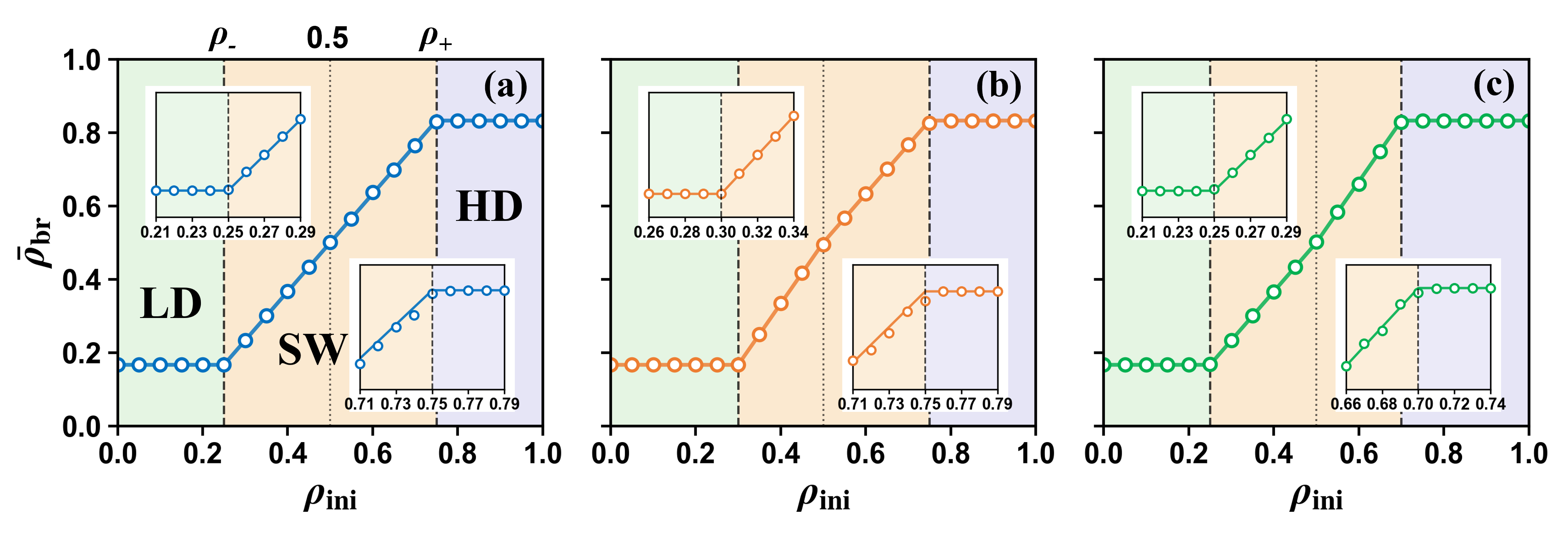}
\caption{Phase diagram of the branches of the $1$-$3$-$1$
model. The simulation results (circles) and theoretical
estimates (curves) of $\bar{\rho}_{\rm br}$ are plotted for $(L_1, L_2, L_3)={\rm (a)\ } (200, 200, 200), {\rm (b)\ } (400, 200, 200), {\rm and \ (c)\ } (200, 200, 400)$, respectively. The dashed (dotted) lines indicate $\rho_{\rm ini}=\rho_-$ and $\rho_{\rm ini}=\rho_+$ ($\rho_{\rm ini}=0.5$). The insets show zoomed views around $\rho_{\rm ini}=\rho_-$ and $\rho_{\rm ini}=\rho_+$.}
\label{fig:comparison_1n1_phase}
\end{figure*}

Based on the above analytical discussions, the phase of branches depends on $\rho_{\rm ini}$ with fixed $(n,L_1,L_2,L_3)$.  In the main text, we
focus on the $1$-$3$-$1$ model, so that the results can be compared directly with those of the $0$-$3$-$0$ model discussed below.  Figure~\ref{fig:comparison_1n1_phase} presents the phase
diagram for the $1$-$3$-$1$ model, comparing the simulation results and
the theoretical estimates of $\bar{\rho}_{\rm br}$, given by Eq. (\ref{eq:barrhobr_1n1}), for various sets of $(L_1, L_2, L_3)$ with $n=3$. The three shaded regions, divided by $\rho_{\rm ini}=\rho_-$ and $\rho_{\rm ini}=\rho_+$, represent the LD (pale green), SW (light beige), and HD (light lavender) phases.

The simulation results of $\bar{\rho}_{\rm br}$ exhibit excellent agreement with the theoretical results. In particular, the plateau regions in the LD
and HD phases and the monotonic variation in the SW phase are well reproduced. Additional results for other sets of \((n,L_1,L_2,L_3)\) are summarized
in Appendix~\ref{sec:pd_1n1_app}, where we confirm the same behavior. 

We next examine more explicitly how the phase boundaries depend on the
geometry.  The analytical estimates predict that the lower boundary
$\rho_-$ is controlled by $L_1$, whereas the upper boundary $\rho_+$ is controlled by $L_3$.  To test this prediction quantitatively, we regard $\rho_-$ and $\rho_+$ as fitting parameters and determine them by least-squares fitting of
$\bar{\rho}_{\rm br}$ to Eq.~\eqref{eq:barrhobr_1n1}.  The details of the fitting procedure are described in Appendix~\ref{app:fitting_rhopm}. Figure~\ref{fig:comparison_rhopm_geometry} presents the geometry dependence of (a) \(\rho_-\) and (b) \(\rho_+\), comparing the numerical estimates with the theoretical predictions from Eqs. (\ref{eq:rho_minus_1n1}) and
(\ref{eq:rho_plus_1n1}), respectively, for $n\in\{2,3,5\}$.
The simulation results of \(\rho_-\) and \(\rho_+\) exhibit excellent agreement with the theoretical results. The agreement in Figs.~\ref{fig:comparison_rhopm_geometry}
therefore provides the geometric prediction:
$\rho_-$ is controlled by $L_1$, independently of $L_3$, whereas $\rho_+$ is controlled by $L_3$, independently of $L_1$.

Having established the global phase behavior, we next examine the spatial
density profiles to mainly clarify how the SW phase is realized locally.

\begin{figure}[t]
\centering
\includegraphics[width=\columnwidth]{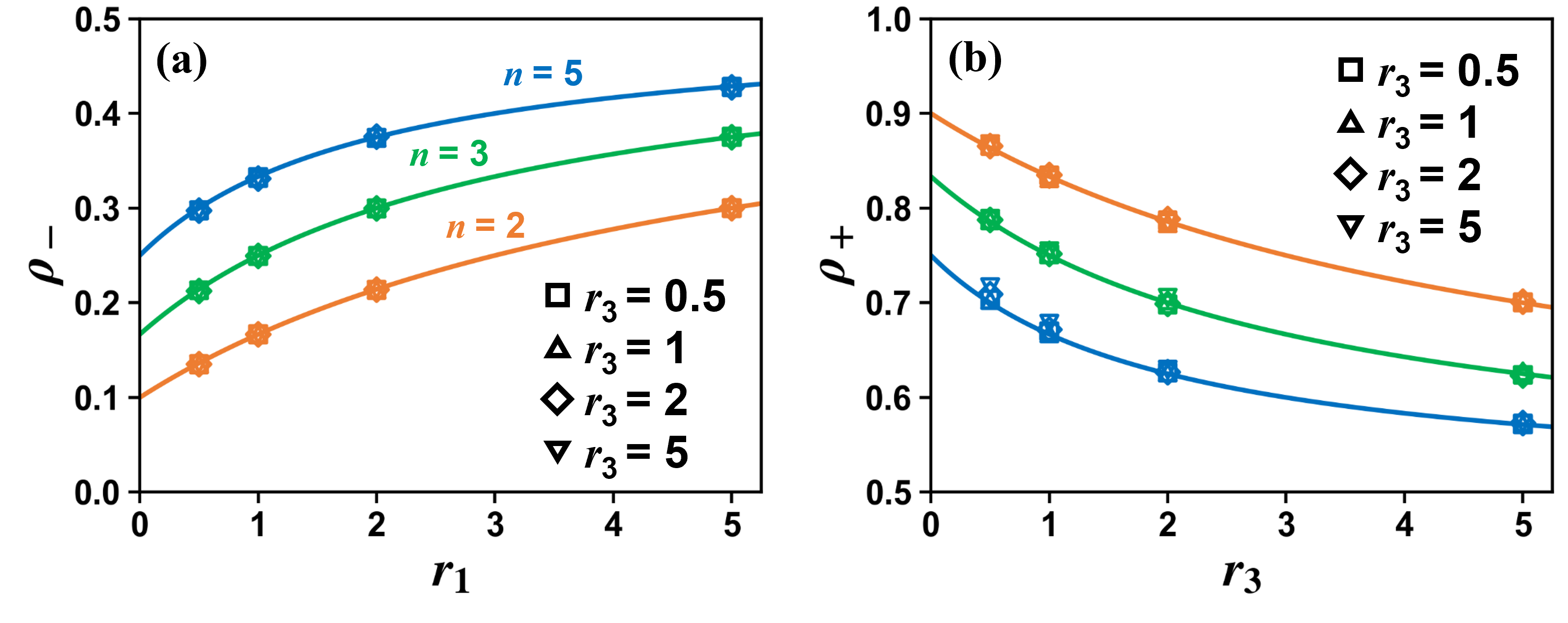}
\caption{Geometry dependence of the SW phase boundaries, i.e., $\rho_-$ and $\rho_+$, in the
$1$-$n$-$1$ model with $n\in\{2, 3, 5\}$. (a) Numerical results (symbols) and the theoretical estimates (curves), i.e., Eq. (\ref{eq:rho_minus_1n1}), of $\rho_-$ as a function of $r_1$. The squares, triangles, diamonds, and reversed triangles represent the simulation results of $r_3\in\{0.5, 1, 2, 5\}$, respectively. (b) Simulation results (symbols) and the theoretical estimates (curves), i.e., Eq. (\ref{eq:rho_plus_1n1}), of $\rho_+$ as a function of $r_3$. The squares, triangles, diamonds, and reversed triangles represent the simulation results of $r_1\in\{0.5, 1, 2, 5\}$, respectively. For each parameter value, the simulation symbols overlap with each other in both panels.}
\label{fig:comparison_rhopm_geometry}
\end{figure}

\subsection{Density Profiles and Local Shock-Wave Picture}
\label{sec:DP1n1}

We then investigate the branch density profile $\rho_{\rm br}(x)$. 
In the LD and HD phases, $\rho_{\rm br}$ does not depend on $x$, i.e., $\rho_{\rm br}(x)=\rho_{\rm L}$ and $\rho_{\rm br}(x)=\rho_{\rm H}$, respectively.

Then, we consider $\rho_{\rm br}(x)$ in the SW phase. As a coarse-grained approximation, we regard the time-averaged branch profile as a coexistence of an LD-like region and an HD-like region in the SW phase.
Let $l$ be the effective fraction of the branch occupied by the LD-like region. Particle-number conservation in the branch gives
\begin{equation}
\rho_{\rm L}l
+
\rho_{\rm H}
\left(
1-l
\right)
=
\bar{\rho}_{\rm br}.
\label{eq:lever_rule_1n1}
\end{equation}
Therefore,
\begin{equation}
l
=
\frac{
\rho_{\rm H}
-
\bar{\rho}_{\rm br}
}
{
\rho_{\rm H}
-
\rho_{\rm L}
},
\qquad
h
=
1-l
=
\frac{
\bar{\rho}_{\rm br}
-
\rho_{\rm L}
}
{
\rho_{\rm H}
-
\rho_{\rm L}
},
\label{eq:h_1n1_general}
\end{equation}
where $h$ is the effective fraction occupied by the HD-like component.

\begin{figure}[t]
\centering
\includegraphics[width=0.6\columnwidth]{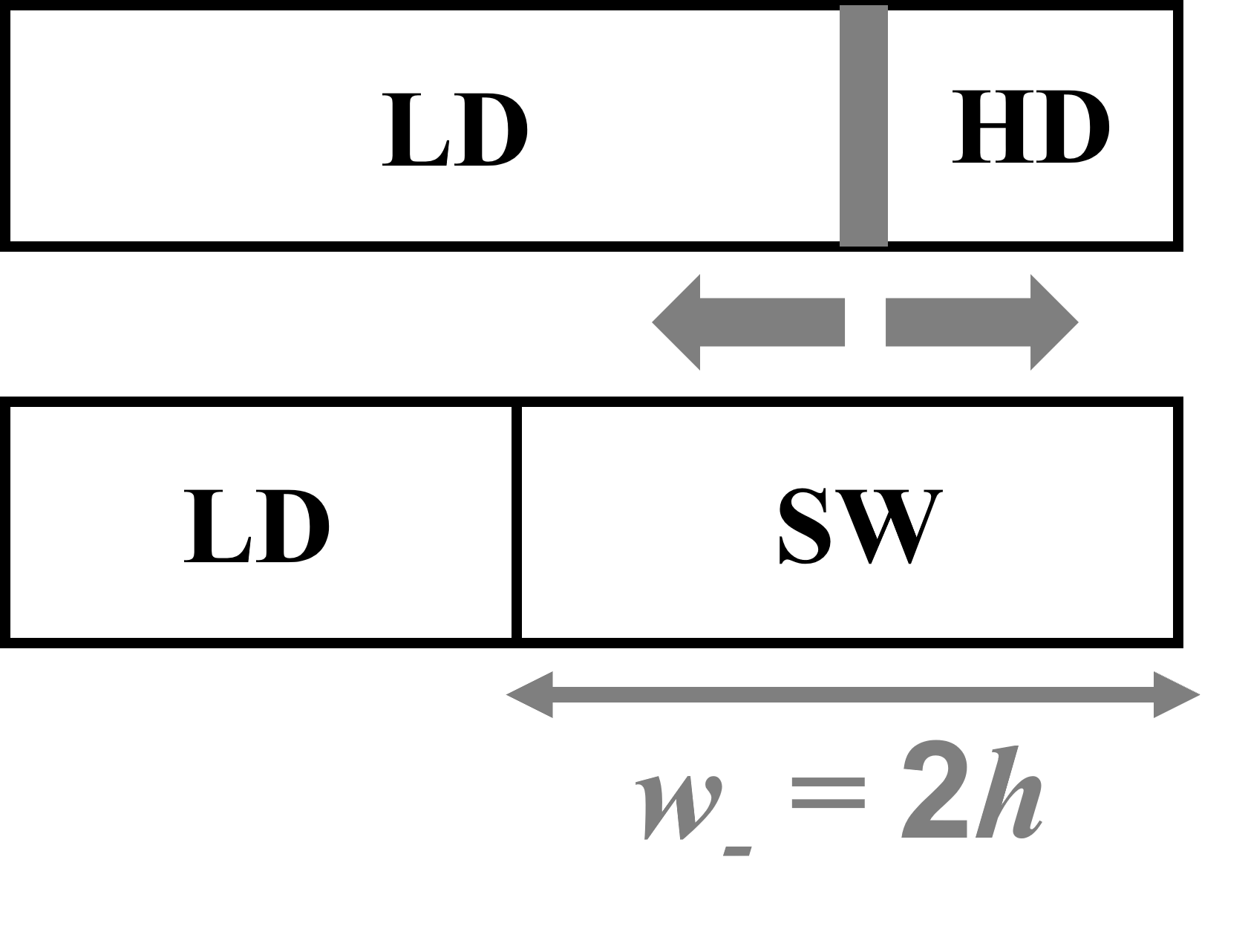}
\caption{Schematic illustration of the SW phase.  If the boundary between
the LD-like and HD-like regions were fixed, the HD-like component would
occupy an effective fraction \(h\).  When the boundary fluctuates in
time, the time-averaged profile forms a coexistence region whose width
is estimated as \(w_-=2h\).}
\label{fig:hw}
\end{figure}

The relation between the effective HD-like component and the SW phase is illustrated schematically in Fig.~\ref{fig:hw}.
If the shock were fixed, the same excess particle number would correspond
to an HD-like region of length \(h\).  In the time-averaged profile,
however, the shock position fluctuates over a finite interval, producing
a linear ramp between \(\rho_{\rm L}\) and \(\rho_{\rm H}\). For $\rho_-\leq\rho_{\rm ini}<1/2$, since the
average height of this ramp above the LD background is only a half of
\(\rho_{\rm H}-\rho_{\rm L}\), the width of the ramp $w_-$ must be twice the
effective HD fraction, i.e.,
\begin{equation}
w_-
=
2h
=
\frac{
2
\left(
\bar{\rho}_{\rm br}^{(-)}
-
\rho_{\rm L}
\right)
}
{
\rho_{\rm H}
-
\rho_{\rm L}
}.
\label{eq:w_minus_1n1}
\end{equation}

This gives $w_-=0$ at $\rho_{\rm ini}=\rho_-$ and $w_-=1$ at $\rho_{\rm ini}=1/2$. The branch density profile ${\rho}_{\rm br}(x)$ for $\rho_-\leq\rho_{\rm ini}<1/2$ is then approximated as
\begin{equation}
\rho_{\rm br}(x)
=
\rho_{\rm L}
+
\frac{\rho_{\rm H}-\rho_{\rm L}}{w_-}
\max\left\{0,\,x-\left(1-w_-\right)\right\}.
\label{eq:avg_profile_1n1_lower}
\end{equation}

For \(1/2 \le \rho_{\rm ini}<\rho_+\), the same construction is applied to holes.
The width of the local low-density component is
\begin{equation}
w_+
=
\frac{
2
\left(
\bar{\rho}_{{\rm h},{\rm br}}^{(+)}
-
\rho_{\rm L}
\right)
}
{
\rho_{\rm H}-\rho_{\rm L}
}.
\label{eq:w_plus_1n1}
\end{equation}
This gives $w_+=1$ at $\rho_{\rm ini}=1/2$ and $w_+=0$ at
$\rho_{\rm ini}=\rho_+$. ${\rho}_{\rm br}(x)$ is approximated as
\begin{equation}
{\rho}_{\rm br}(x)
=
\rho_{\rm H}
+
\dfrac{\rho_{\rm H}-\rho_{\rm L}}{w_+}\min\left(0,\,x-w_+\right).
\label{eq:avg_profile_1n1_upper}
\end{equation}
Eventually, $\rho_{\rm br}(x)$ is summarized as
\begin{widetext}
\begin{equation}
\rho_{\rm br}(x)=
\begin{cases}
\rho_{\rm L}
&{\rm for \ } 0\leq\rho_{\rm ini}<\rho_-,\\
\rho_{\rm L}
+
\dfrac{\rho_{\rm H}-\rho_{\rm L}}{w_-}
\max\left\{0,\,x-\left(1-w_-\right)\right\}
&{\rm for \ } \rho_-\leq\rho_{\rm ini}<\dfrac{1}{2},\\
\rho_{\rm H}
+
\dfrac{\rho_{\rm H}-\rho_{\rm L}}{w_+}\min\left(0,\,x-w_+\right)
&{\rm for \ } \dfrac{1}{2}\leq\rho_{\rm ini}<\rho_+,\\
\rho_{\rm H}
&{\rm for \ } \rho_+<\rho_{\rm ini}\leq1.
\end{cases}
\label{eq:rhobr_1n1}
\end{equation}
\end{widetext}
We stress that the relation
\begin{equation}
\bar{\rho}_{\rm br}=\int_0^1 \rho_{\rm br}(x) dx
\label{eq:integral}
\end{equation}
holds.

\begin{figure*}[t]
\centering
\includegraphics[width=\textwidth]{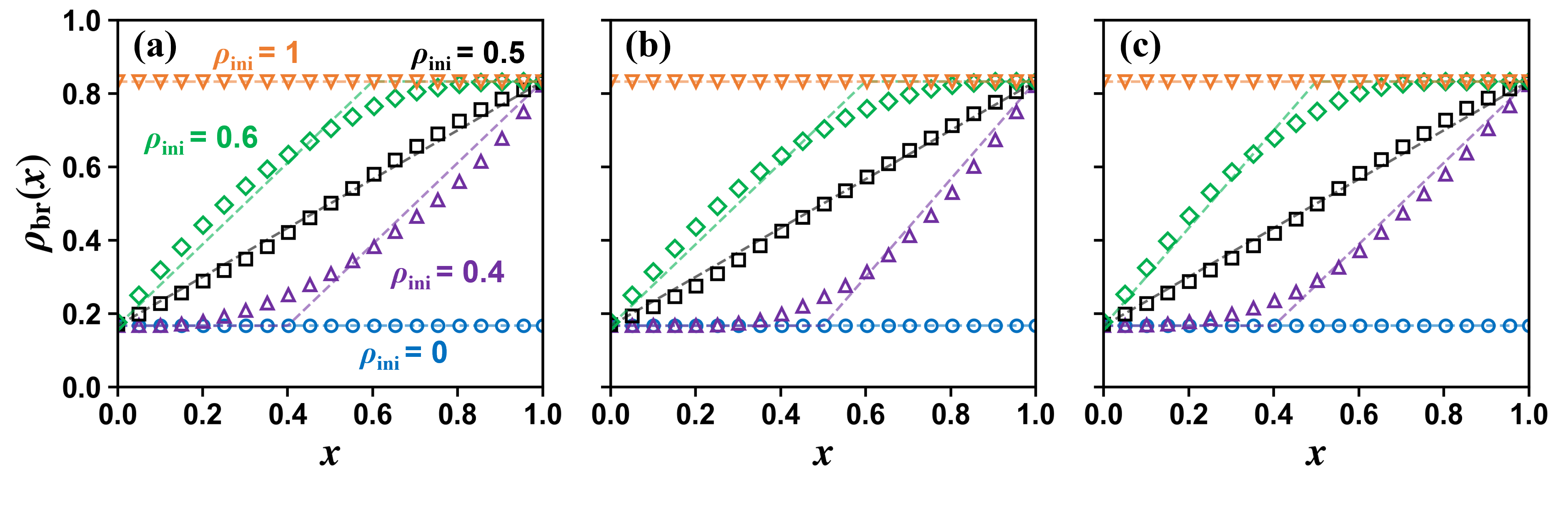}
\caption{Branch density profiles of the $1$-$3$-$1$
model for $(L_1, L_2, L_3)={\rm (a)\ } (200, 200, 200), {\rm (b)\ } (400, 200, 200), {\rm and \ (c)\ } (200, 200, 400)$, respectively. The blue, purple, black, green, and orange symbols (dashed curves) represent the simulation results (theoretical curves) of $\rho_{\rm br}(x)$ for $\rho_{\rm ini}\in\{0, 0.4, 0.5, 0.6, 1\}$.}
\label{fig:comparison_1n1_profiles}
\end{figure*}

Figure~\ref{fig:comparison_1n1_profiles} presents the density profile of the branches, comparing the simulation results with the theoretical prediction for ${\rho}_{\rm br}(x)$, given by Eq. (\ref{eq:rhobr_1n1}), with the same sets of $(n, L_1, L_2, L_3)$ as Fig.~\ref{fig:comparison_1n1_phase}. In all the panels, the system is theoretically expected to yield to the LD phase with $\rho_{\rm ini}=0$, the SW phase with $\rho_{\rm ini}\in\{0.4, 0.5, 0.6\}$, and the HD phase with $\rho_{\rm ini}=1$, respectively.

The simulation results show good agreement with the theoretical predictions. As expected, the profiles are flat in the LD and HD phases, whereas tilted profiles appear in the SW phase.  This behavior supports the local shock-wave picture: the tilted profile is interpreted as the time average of local
shocks separating LD-like and HD-like regions in the branches. Compared with $\bar{\rho}_{\rm br}$, the deviations from the theory are more visible in
${\rho}_{\rm br}(x)$, which is more sensitive to branch-to-branch fluctuations.

\begin{figure*}[t]
\centering
\includegraphics[width=\textwidth]{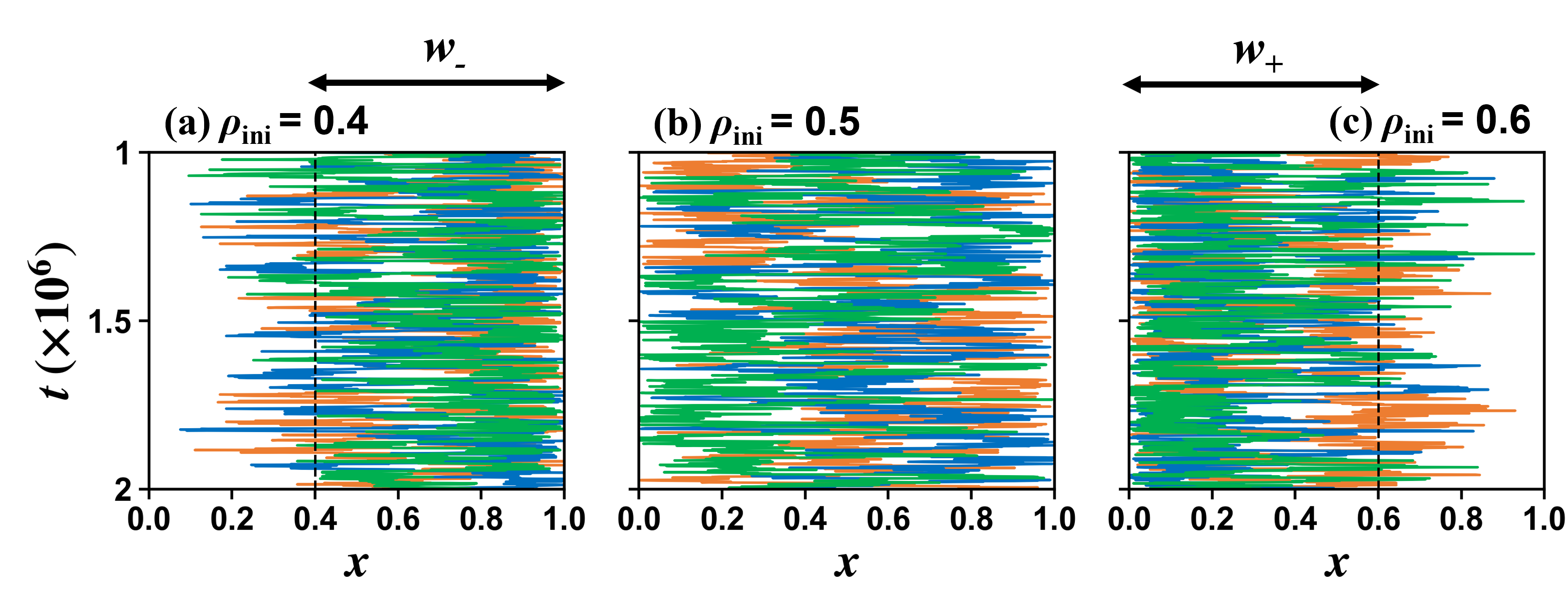}
\caption{Trajectories of the second-class particles for three branches (blue, green, and orange) of $1$-$3$-$1$ model for (a) $\rho_{\rm ini}=0.4$, (b) $\rho_{\rm ini}=0.5$, and (c) $\rho_{\rm ini}=0.6$, respectively, with $(L_1, L_2, L_3)=(200, 200, 200)$. The value $t$ represents the simulation time.}
\label{fig:second_class_1n1}
\end{figure*}

To examine the instantaneous motion of the local shocks, we next track second-class particles. Following the standard
use of second-class particles as microscopic probes of shock positions in
exclusion processes~\cite{ferrari1992shock,derrida1993microscopic}, we introduce them as passive markers of the interfaces between LD-like and HD-like regions.  Figure~\ref{fig:second_class_1n1} shows the trajectories obtained in a
single trial. For $\rho_{\rm ini}=0.4$, the trajectories are biased toward the downstream side, whereas for $\rho_{\rm ini}=0.6$, they are
biased toward the upstream side.  For $\rho_{\rm ini}=0.5$, they explore almost the whole branch.  

The observed trajectories are broadly consistent with the theoretical local shock-wave picture; however, they are not sharply confined to the estimated ranges $w_-$ and $w_+$.  In particular, the instantaneous
positions spread over a wider region than the simple coarse-grained estimate suggests.  Thus, the theory should be understood as describing the average accessible region of the shock, rather than the precise
trajectory in a single realization.  The tilted density profiles in Fig.~\ref{fig:comparison_1n1_profiles} are therefore interpreted as time-averaged signatures of these fluctuating local shocks.

We note that for $(\alpha,\beta)\ne(1,1)$, the density profile is controlled by $(\alpha, \beta)$ rather than by $\rho_{\rm ini}$, so that the initial-density dependence does not persist. Details are given in Appendix~\ref{sec:alphabetaneq1}.

\section{Analyses of $0$-$n$-$0$ model}
\label{sec:analyses0n0}

We next consider the \(0\)-\(n\)-\(0\) model to clarify how the boundary
geometry affects the branch dynamics. The line of argument is the same as
for the \(1\)-\(n\)-\(1\) model, but the absence of the upstream and downstream
single-lane segments changes the particle-number constraint and hence the
phase boundaries.

\subsection{Phase Diagram and Mean Branch Density}
\label{sec:PD0n0}
The $0$-$n$-$0$ model consists only of $n$ parallel branches of length $L$. There is no upstream lane before the branches and no downstream lane after them. Particles are injected directly into one of the branches and leave directly from the right boundary of the branches.

The first newly injected particle reaches the exit boundary after a time
of order
\begin{equation}
t_{\rm arr}
\simeq
L .
\label{eq:t_arr_0n0}
\end{equation}
The number of particles initially existing in the branches is
\begin{equation}
N_{\rm ini}
=
n\rho_{\rm ini}L.
\label{eq:N_ini_0n0}
\end{equation}

We next estimate the number of particles that can leave the system
before the first newly injected particle reaches the exit. The early-time exit flow depends on the initial
density. For low initial density, the total exit flow is proportional to
the number of branches and to $\rho_{\rm ini}$. However, because at most
one particle can leave the system in one time step, the total exit flow
is bounded by unity. Thus,
\begin{equation}
Q_{\rm ini,ex}
=
\begin{cases}
n\rho_{\rm ini}
& {\rm for \ } 0\leq\rho_{\rm ini}<\dfrac{1}{n},\\
1
& {\rm for \ } \dfrac{1}{n}\leq\rho_{\rm ini}\leq\dfrac{1}{2}.
\end{cases}
\label{eq:Q_ini_ex_0n0}
\end{equation}
Therefore, the number of particles that can leave the system during the time interval $t_{\rm arr}$ is
\begin{equation}
N_{\rm ex}
\simeq
Q_{\rm ini,ex}t_{\rm arr},
\label{eq:N_ex_0n0_def}
\end{equation}
or explicitly,
\begin{equation}
N_{\rm ex}
=
\begin{cases}
n\rho_{\rm ini}L
& {\rm for \ } 0\leq\rho_{\rm ini}<\dfrac{1}{n},\\
L
& {\rm for \ } \dfrac{1}{n}\leq\rho_{\rm ini}\leq\dfrac{1}{2}.
\end{cases}
\label{eq:N_ex_0n0}
\end{equation}
For $0\leq\rho_{\rm ini}<1/n$, $N_{\rm ini}=N_{\rm ex}$, indicating that the initially existing particles are removed from the system by the time when the newly injected particles catch up. Thus, the system
relaxes to the LD state in this region.
For $1/n\leq\rho_{\rm ini}\leq1/2$, we have
\begin{align}
N_{\rm ini}-N_{\rm ex}
&=
n\rho_{\rm ini}L-L
\nonumber\\
&=
L
\left(
n\rho_{\rm ini}-1
\right)
>0.
\label{eq:excess_particles_0n0}
\end{align}
Thus, a macroscopic number of particles remains when the newly injected
particles reach the exit, and a local coexistence layer is formed. The
lower critical initial density is therefore
\begin{equation}
\rho_-
=
\frac{1}{n}.
\label{eq:rho_minus_0n0}
\end{equation}
Because the $0$-$n$-$0$ geometry has no additional upstream or
downstream lane, the particle-hole counterpart is obtained simply as
\begin{equation}
\rho_+
=
1-\frac{1}{n}.
\label{eq:rho_plus_0n0}
\end{equation}
Thus, the expected phases are summarized as
\begin{equation}
\begin{cases}
{\rm LD}
& {\rm for \ } 0\leq\rho_{\rm ini}<\rho_-,\\
{\rm SW}
& {\rm for \ } \rho_-\leq\rho_{\rm ini}\leq\rho_+,\\
{\rm HD}
& {\rm for \ } \rho_+<\rho_{\rm ini}\leq1.
\end{cases}
\label{eq:phase_0n0}
\end{equation}

It is worth noting that the $0$-$n$-$0$ model is not obtained by simply
taking the limit $L_1\to0$ in the $1$-$n$-$1$ result. From
Eq.~\eqref{eq:rho_minus_1n1}, this limit gives
\begin{equation}
\rho_-
\to
\frac{1}{2n},
\label{eq:rho_cr_1n1_L1_zero}
\end{equation}
whereas the present $0$-$n$-$0$ model gives $\rho_-=1/n$. This discrepancy reflects the difference in the boundary rule.  In the
\(1\)-\(n\)-\(1\) model, the \(L_1\to0\) limit still contains a
single-lane bottleneck at the entrance to the branches, so the maximal
injection current is \(1/2\).  Hence this limit is not equivalent to the
\(0\)-\(n\)-\(0\) model, where particles can be injected directly into
the \(n\) branches.  The effective injection density is therefore
\(\rho_-=1/(2n)\), consistent with Eq.~\eqref{eq:rho_cr_1n1_L1_zero}.

We next estimate the branch-averaged density $\bar{\rho}_{\rm br}$. The values of $\bar{\rho}_{\rm br}$ in the LD and HD phases in each branch are first considered.  In the LD and HD phases, the branch density is spatially uniform in the bulk and is determined by the branching flow in the LD phase and by the merging flow in the HD phase. The total flow is bounded by the maximal-flow value $1$ and is shared by the $n$ branches; hence $\bar{\rho}_{\rm br}$ are estimated as
\begin{equation}
\bar{\rho}_{\rm br}=
\begin{cases}
\dfrac{1}{n}(=\rho_{\rm L})
& {\rm for \ LD},\\
1-\dfrac{1}{n}(=\rho_{\rm H})
& {\rm for \ HD}.
\end{cases}
\label{eq:rho_LH_0n0}
\end{equation}

Then, we consider \(\bar{\rho}_{\rm br}\) in the SW phase by focusing on
the particle balance.  Since the catching-up time is given by
Eq.~\eqref{eq:t_arr_0n0}, about \(t_{\rm arr}\) particles enter the
system during this time.  This number is equal to \(N_{\rm ex}\), so that
the inflow and outflow cancel in the estimate of the branch particle
number.  Hence, the number of particles remaining in the branches is
estimated as
\begin{equation}
N_{\rm br}
=
\rho_{\rm ini} nL,
\end{equation}
and therefore the corresponding branch-averaged density is
\begin{equation}
\bar{\rho}_{\rm br}
=
\rho_{\rm ini}.
\end{equation}
Eventually, the theoretical branch-averaged density $\bar{\rho}_{\rm br}$ is summarized as
\begin{equation}
\bar{\rho}_{\rm br}=
\begin{cases}
\dfrac{1}{n}
&{\rm for \ } 0\leq\rho_{\rm ini}<\rho_-,\\
\rho_{\rm ini}
&{\rm for \ } \rho_-\leq\rho_{\rm ini}<\rho_+,\\
1-\dfrac{1}{n}
&{\rm for \ } \rho_+<\rho_{\rm ini}\leq1.
\end{cases}
\label{eq:rhobr_0n0}
\end{equation}
We note that $\bar{\rho}_{\rm br}=1/2$ for $\rho_{\rm ini}\in[0,1]$ with $n=2$. This can be understood intuitively from the fact that for $n=2$ one particle is injected into one of the two empty entrances at each step and one particle can leave from one of the two occupied exits in an alternating manner.

\begin{figure}[t]
\centering
\includegraphics[width=\columnwidth]{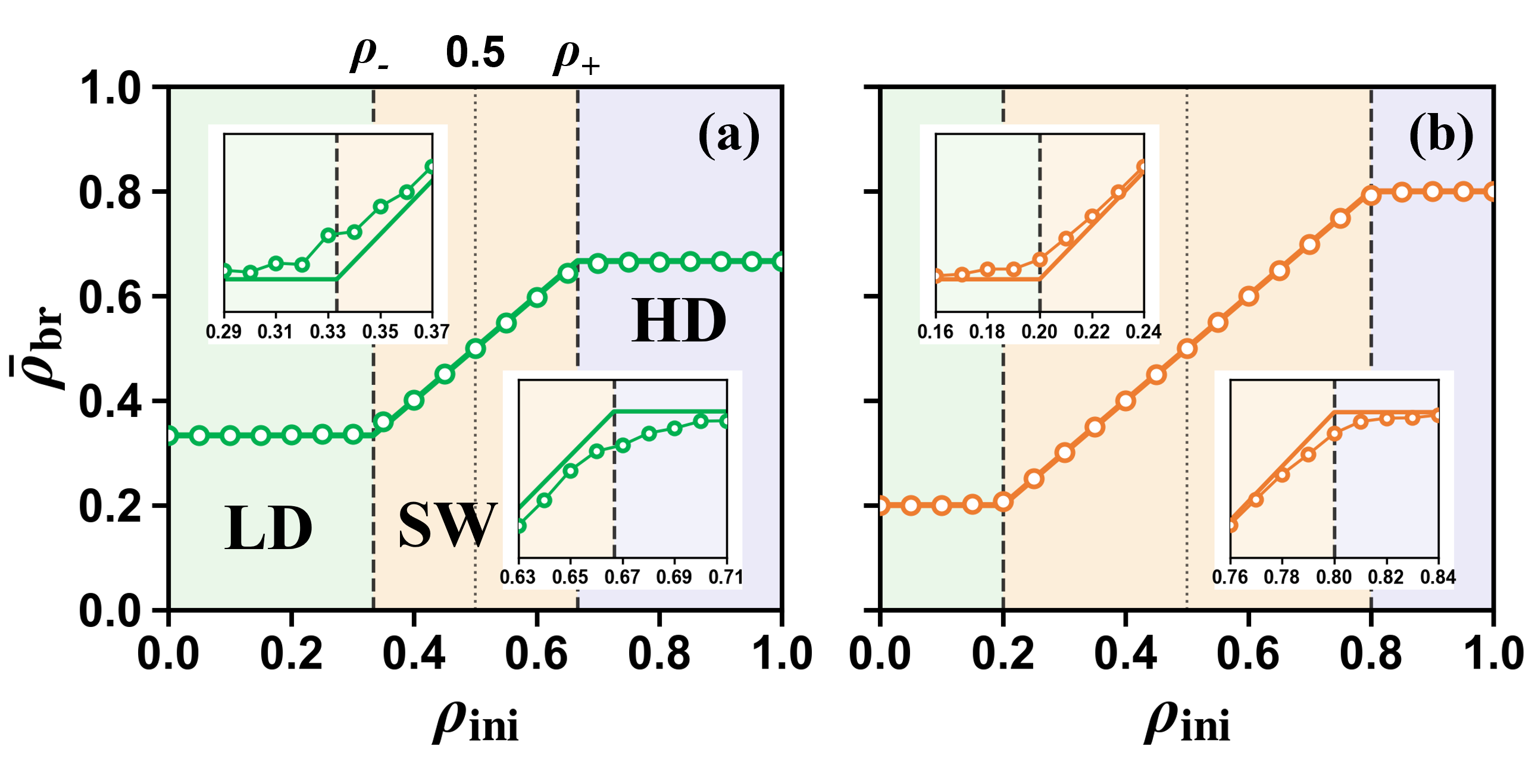}
\caption{Phase diagram of the branches of the $0$-$3$-$0$
model. The simulation results (circles) and theoretical
estimates (curves) of $\bar{\rho}_{\rm br}$ are plotted for (a) $n=3$ and (b) $n=5$ with $L=200$. The dashed (dotted) lines indicate $\rho_{\rm ini}=\rho_-$ and $\rho_{\rm ini}=\rho_+$ ($\rho_{\rm ini}=0.5$). The insets show zoomed views around $\rho_{\rm ini}=\rho_-$ and $\rho_{\rm ini}=\rho_+$.}
\label{fig:comparison_0n0_phase}
\end{figure}

The phase of branches depends on $\rho_{\rm ini}$ with fixed $n$, independently of $L$. Figure~\ref{fig:comparison_0n0_phase} presents the phase diagram along \(\rho_{\rm ini}\) for the branches
of the $0$-$n$-$0$ model, comparing the simulation results with the theoretical prediction for $\bar{\rho}_{\rm br}$, given by Eqs. (\ref{eq:rhobr_0n0}), for $n\in\{3,5\}$ with $L=200$. The three shaded regions, divided by $\rho_{\rm ini}=\rho_-$ and $\rho_{\rm ini}=\rho_+$, represent the LD (pale green), SW (light beige), and HD (light lavender) phases.

Like the $1$-$n$-$1$ model, the simulation results of $\bar{\rho}_{\rm br}$ exhibit excellent agreement with the theoretical results.

We next examine more explicitly how the phase boundaries depend on $n$. 
We regard $\rho_-$ and $\rho_+$ as fitting parameters and determine them by least-squares fitting of
$\bar{\rho}_{\rm br}$ to Eq.~\eqref{eq:rhobr_0n0}.  The details of the fitting procedure are described in Appendix~\ref{app:fitting_rhopm}. Figure~\ref{fig:comparison_rhopm_0n0} shows the $n$-dependence of $(\rho_-, \rho_+)$, comparing the numerical estimates with the theoretical predictions from Eqs. (\ref{eq:rho_minus_0n0}) and
(\ref{eq:rho_plus_0n0}), respectively. The simulation results of \(\rho_-\) and \(\rho_+\) exhibit excellent agreement with the theoretical results.

Having established the global phase behavior, we next examine the spatial
density profiles to clarify how the SW phase is realized locally.

\begin{figure}[t]
\centering
\includegraphics[width=0.8\columnwidth]{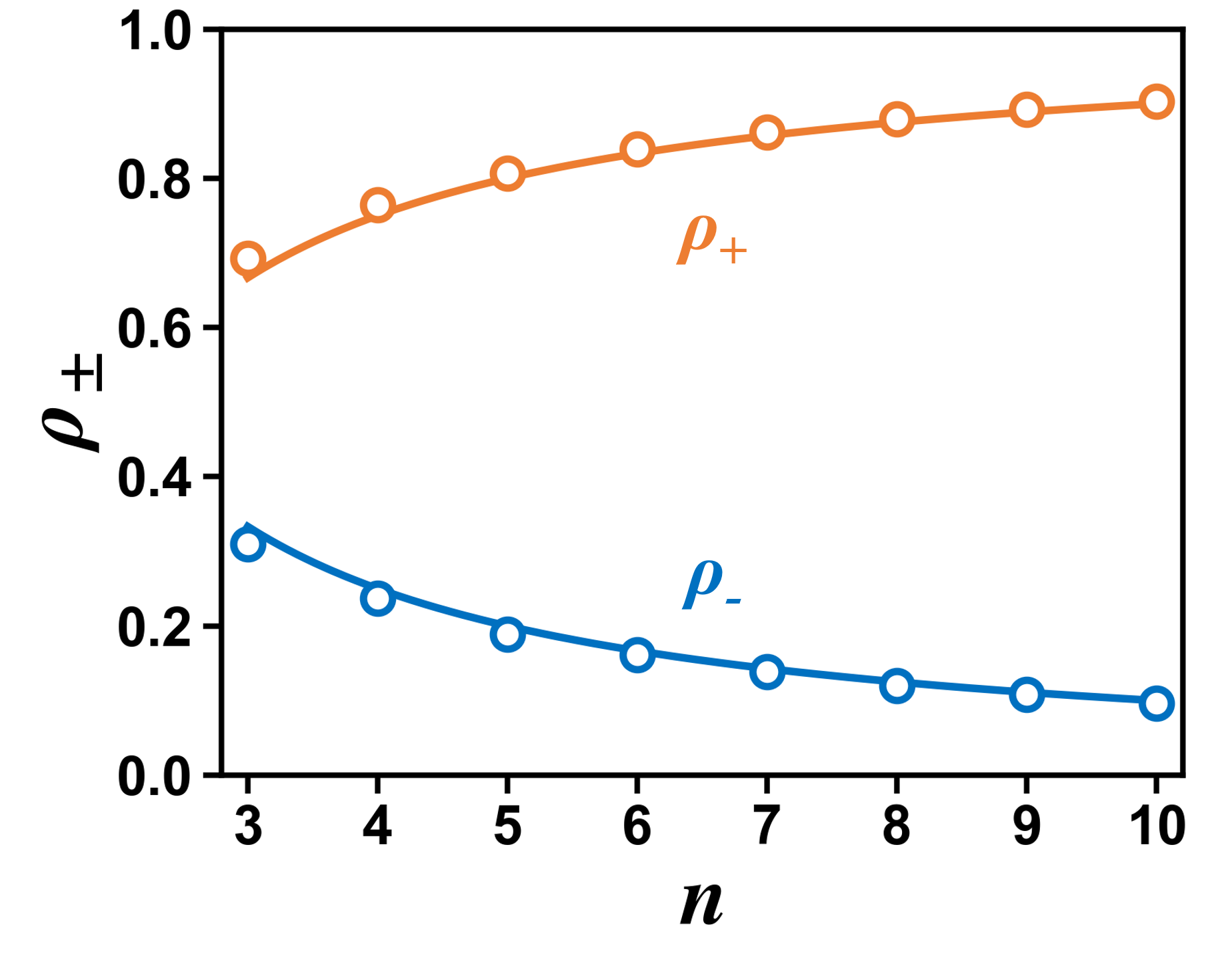}
\caption{$n$-dependence of the SW phase boundaries, i.e., $\rho_-$ and $\rho_+$, in the $0$-$n$-$0$ model. The blue and orange circles (curves) represent the numerical results (theoretical predictions) of $\rho_-$ and $\rho_+$, respectively.}
\label{fig:comparison_rhopm_0n0}
\end{figure}

\subsection{Density Profiles and Local Shock-Wave Picture}
\label{sec:DP0n0}

We then investigate the branch density profile $\rho_{\rm br}(x)$. The construction is the same as that used for the \(1\)-\(n\)-\(1\) model; however, the LD- and HD-like plateau densities are
\(\rho_{\rm L}=1/n\) and \(\rho_{\rm H}=1-1/n\), whereas in the SW phase
\(\bar{\rho}_{\rm br}=\rho_{\rm ini}\) for the \(0\)-\(n\)-\(0\) model.
Thus, for \(\rho_-\le \rho_{\rm ini}<1/2\), the width of the local
coexistence region is
\begin{equation}
w_-
=
\frac{
2\left(\bar{\rho}_{\rm br}-\rho_{\rm L}\right)
}{
\rho_{\rm H}-\rho_{\rm L}
}.
\label{eq:w_minus_0n0}
\end{equation}
This gives $w_-=0$ at $\rho_{\rm ini}=\rho_-$ and $w_-=1$ at $\rho_{\rm ini}=1/2$. The branch density profile ${\rho}_{\rm br}(x)$ for $\rho_-\leq\rho_{\rm ini}<1/2$ is then approximated as Eq. (\ref{eq:avg_profile_1n1_lower}).
For $1/2\leq\rho_{\rm ini}<\rho_+$, the same construction is applied to holes. With $\rho_{\rm h}=1-\bar{\rho}_{\rm br}$, the width of the local low-density component is
\begin{equation}
w_+
=
\frac{
2\left(\rho_{\rm h}-\rho_{\rm L}\right)
}{\rho_{\rm H}-\rho_{\rm L}
}.
\label{eq:w_plus_0n0}
\end{equation}
This gives $w_+=1$ at $\rho_{\rm ini}=1/2$ and $w_+=0$ at
$\rho_{\rm ini}=\rho_+$. ${\rho}_{\rm br}(x)$ is approximated as Eq. (\ref{eq:avg_profile_1n1_upper}).
Eventually, the theoretical density profile is summarized as the same as Eq. (\ref{eq:rhobr_1n1}).

\begin{figure}[t]
\centering
\includegraphics[width=\columnwidth]{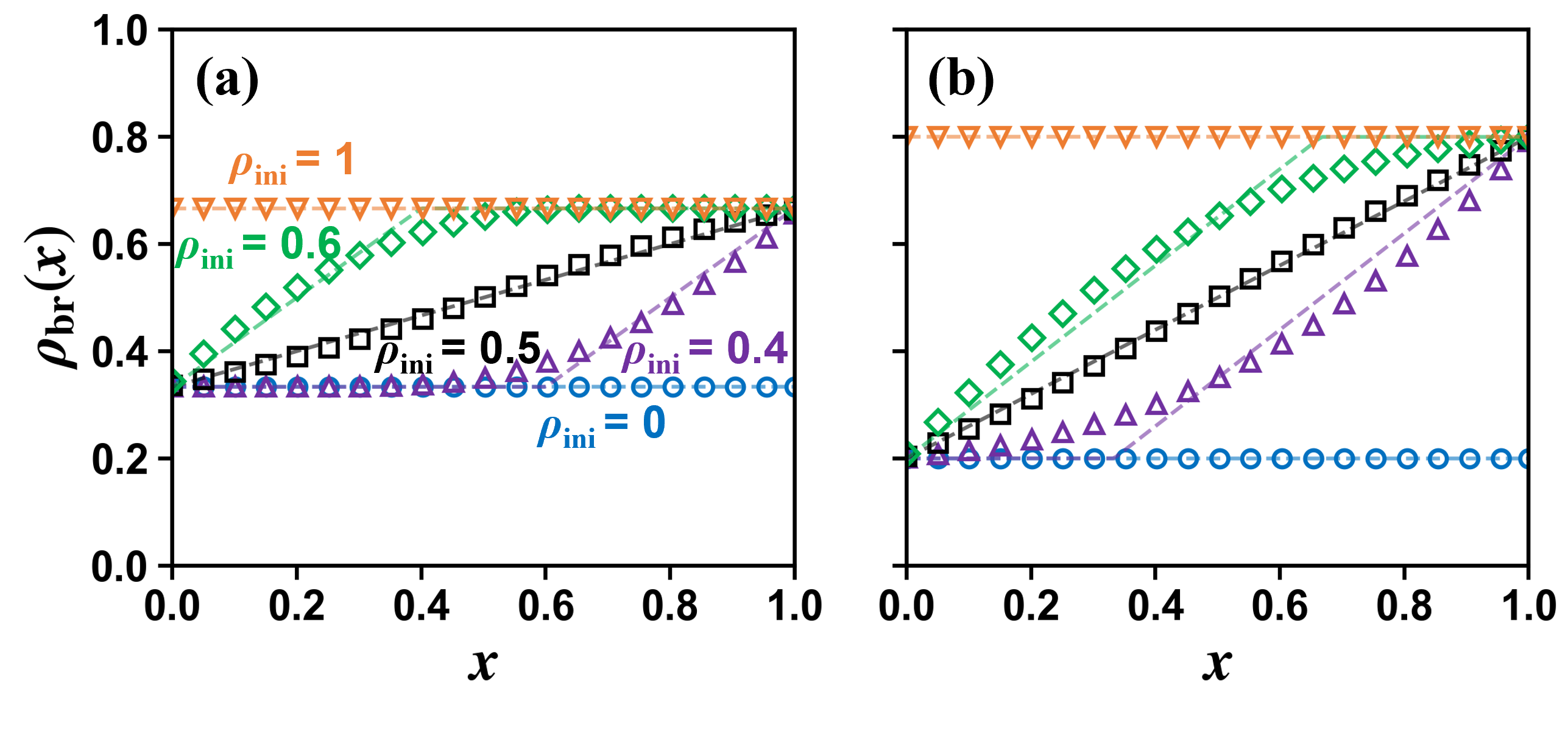}
\caption{Branch density profiles of the $0$-$n$-$0$
model for (a) $n=3$ and (b) $n=5$ with $L=200$. The blue, purple, black, green, and orange symbols (dashed curves) represent the simulation results (theoretical curves) of $\rho_{\rm br}(x)$ for $\rho_{\rm ini}\in\{0, 0.4, 0.5, 0.6, 1\}$.}
\label{fig:comparison_0n0_profiles}
\end{figure}

Figure~\ref{fig:comparison_0n0_profiles} presents the density profile of the branches, comparing the simulation results with the theoretical predictions for ${\rho}_{\rm br}(x)$, also given by Eq.~(\ref{eq:rhobr_1n1}), with the same sets of $(n, L)$ as Fig.~\ref{fig:comparison_0n0_phase}. In both the panels, the system is theoretically expected to yield to the LD phase with $\rho_{\rm ini}=0$, the SW phase with $\rho_{\rm ini}\in\{0.4, 0.5, 0.6\}$, and the HD phase with $\rho_{\rm ini}=1$, respectively.

The simulation results also reproduce the qualitative tendency of the
theoretical density profiles in the $0$-$3$-$0$ model.  The profiles are flat in the LD and HD phases, while tilted profiles appear in the
SW phase.  Thus, the same local shock-wave interpretation can be applied
to the $0$-$n$-$0$ geometry.

\begin{figure*}[t]
\centering
\includegraphics[width=\textwidth]{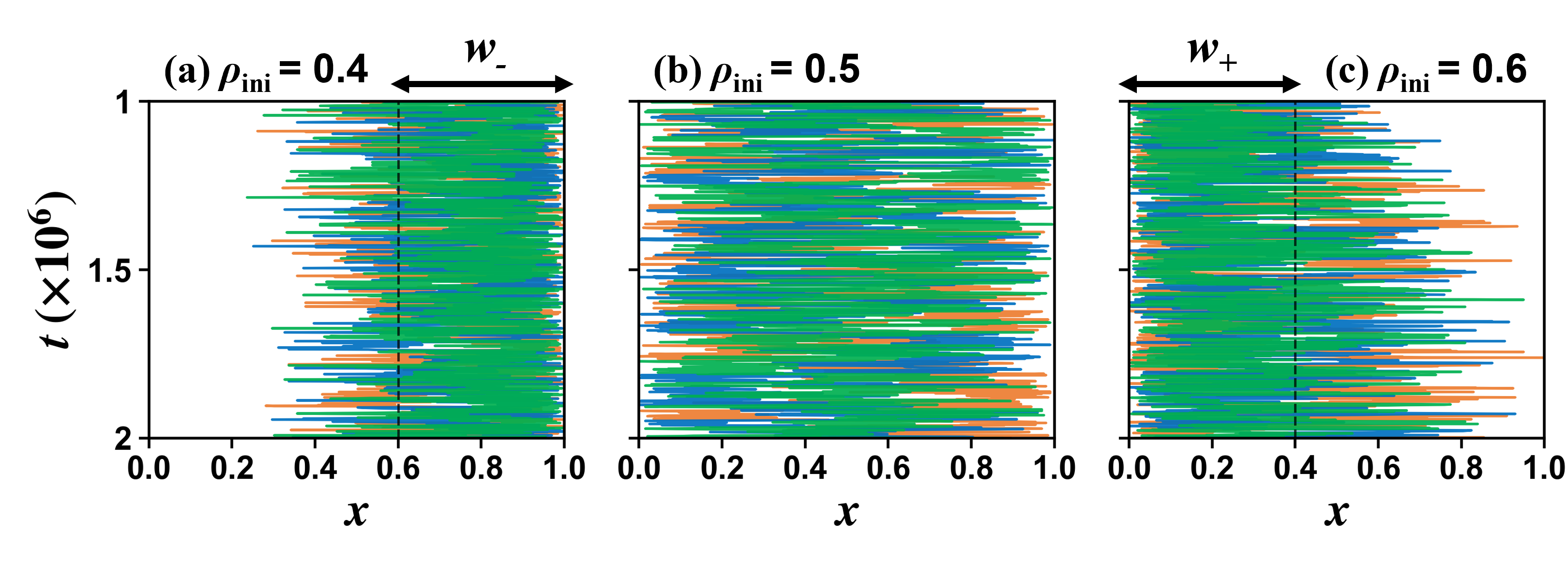}
\caption{Trajectories of the second-class particles for three branches (blue, green, and orange) of $0$-$3$-$0$ model for (a) $\rho_{\rm ini}=0.4$, (b) $\rho_{\rm ini}=0.5$, and (c) $\rho_{\rm ini}=0.6$, respectively, with $L=200$. The dashed lines represent $x=1-w_-$ and $x=w_+$. The value $t$ represents the simulation time.}
\label{fig:second_class_0n0}
\end{figure*}

We again examine the instantaneous interface motion using second-class
particles. Figure~\ref{fig:second_class_0n0} shows the trajectories for
one trial.  As in the $1$-$3$-$1$ model, the trajectories are biased
towards the downstream side for $\rho_{\rm ini}=0.4$, toward the upstream side for $\rho_{\rm ini}=0.6$, and spread over a wide region for $\rho_{\rm ini}=0.5$.  These results support the view that the tilted profiles originate from the time average of moving local shocks. 


We note that the discussions of the density profile for $(\alpha,\beta)\ne(1,1)$ are given in Appendix~\ref{sec:alphabetaneq1} and we obtain the same result as the $1$-$n$-$1$ model.

\section{Conclusion}
\label{sec:conclusion}

In this study, we investigated geometry-controlled,
initial-density-dependent stationary states in branching-merging
exclusion processes.  We considered two boundary geometries for
connecting the parallel branches to the reservoirs: the \(1\)-\(n\)-\(1\)
model and the \(0\)-\(n\)-\(0\) model.  Although these models contain no
explicit friction rule at the merging point, the final state of the
branches can depend on the initial density.

For both models, we derived the critical initial densities separating
the LD, SW, and HD regimes by comparing the number of initially existing
particles with the number of particles that can leave the system before
newly injected particles catch up.  The theoretical estimates showed
good agreement with simulations for the phase diagrams and the
branch-averaged densities.  In the SW phase, the density profiles were
tilted and were interpreted in terms of a local shock-wave picture, in
which LD-like and HD-like regions coexist in the branches.

The comparison between the two boundary geometries clarifies how the
geometry controls the initial-density dependence.  In the
\(1\)-\(n\)-\(1\) geometry, the critical initial densities depend on both
the number of branches and the lengths of the upstream and downstream
single-lane segments.  In contrast, in the \(0\)-\(n\)-\(0\) geometry,
the corresponding estimates depend only on the number of branches.  Thus,
the upstream and downstream single-lane segments do not merely add
extra sites; they change how the initial particle number is converted
into the final branch state.

These results show that initial-condition dependence can arise from the
global geometry of a branching-merging exclusion process, even without
an explicit friction rule at the merging point.  The present models
therefore provide a simple example in which boundary geometry controls
both the critical initial densities and the macroscopic branch state.

\begin{acknowledgments}
This work was supported by JSPS KAKENHI Grant Numbers JP23K21019 and JP26H02177.

The authors used OpenAI Codex (accessed in 2026; version available at
the time of use) to assist with manuscript editing, editing/checking of
author-written analysis codes, and preparation of figures from
author-generated simulation data. The authors independently verified the
scientific content, equations, derivations, simulation data, numerical
results, codes, figures, and final text, and take full responsibility
for the content of the manuscript.
\end{acknowledgments}

\appendix
\section{PHASE DIAGRAMS OF $1$-$n$-$1$ MODEL WITH VARIOUS SETS OF $(n, L_1, L_2, L_3)$}
\label{sec:pd_1n1_app}
\begin{figure*}[t]
\centering
\includegraphics[width=\textwidth]{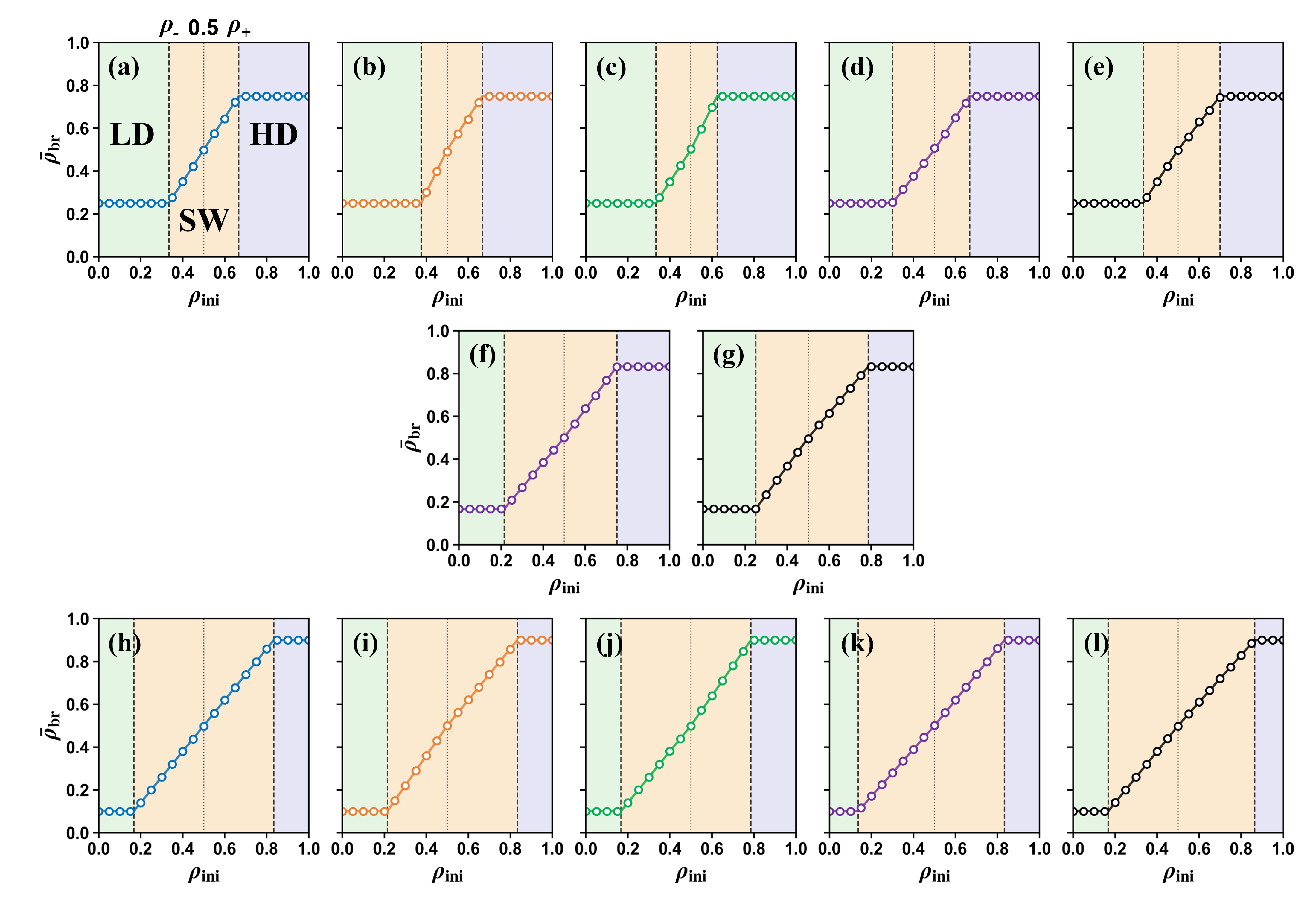}
\caption{Phase diagram of the branches of the \(1\)-\(n\)-\(1\) model.
The simulation results (circles) and theoretical estimates (curves) of
\(\bar{\rho}_{\rm br}\) are plotted as functions of \(\rho_{\rm ini}\).
Panels (a)--(e) show the results for \(n=2\) with
\((L_1,L_2,L_3)=\) (a) \((200,200,200)\), (b) \((400,200,200)\),
(c) \((200,200,400)\), (d) \((100,200,200)\), and
(e) \((200,200,100)\), respectively.  
Panels (f) and (g) show the results for \(n=3\) with
\((L_1,L_2,L_3)=\) (f) \((100,200,200)\) and
(g) \((200,200,100)\), respectively.  
Panels (h)--(l) show the results for \(n=5\) with the same sets of
\((L_1,L_2,L_3)\) as panels (a)--(e), respectively. The dashed (dotted) lines indicate $\rho_{\rm ini}=\rho_-$ and $\rho_{\rm ini}=\rho_+$ ($\rho_{\rm ini}=0.5$).}
\label{fig:comparison_1n1_phase_app}
\end{figure*}

Figure~\ref{fig:comparison_1n1_phase_app} shows additional phase diagrams
of the branches of the \(1\)-\(n\)-\(1\) model for various sets of
\((n,L_1,L_2,L_3)\), complementing Fig.~\ref{fig:comparison_1n1_phase}.
The simulation results are compared with the theoretical estimates of
\(\bar{\rho}_{\rm br}\) given by Eq.~\eqref{eq:barrhobr_1n1}.

For all parameter sets, the simulation results show excellent agreement
with the theoretical curves.  The LD and HD plateaus and the monotonic
variation of \(\bar{\rho}_{\rm br}\) in the SW phase are reproduced in
the same manner as in the main text.  The results also confirm the
geometric dependence of the phase boundaries: changing \(L_1\) shifts the
lower boundary \(\rho_-\), whereas changing \(L_3\) shifts the upper
boundary \(\rho_+\).

\section{BRANCH DENSITY PROFILES OF $1$-$n$-$1$ MODEL WITH VARIOUS SETS OF $(n, L_1, L_2, L_3)$}
\label{sec:bd_1n1_app}

\begin{figure*}[t]
\centering
\includegraphics[width=\textwidth]{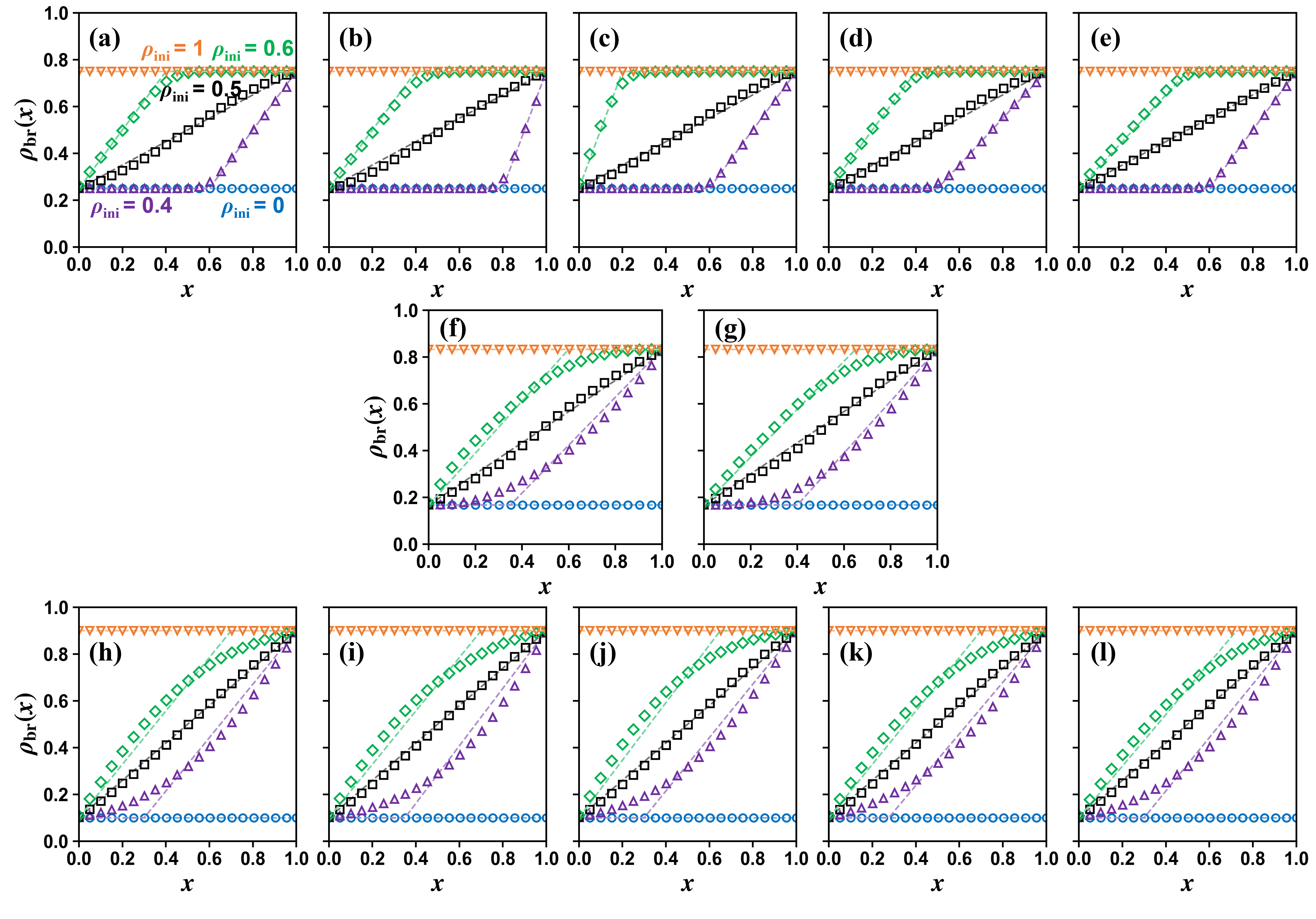}
\caption{Branch density profiles of the \(1\)-\(n\)-\(1\) model. Panels (a)--(e) show the results for \(n=2\) with \((L_1,L_2,L_3)=\) (a) \((200,200,200)\), (b) \((400,200,200)\), (c) \((200,200,400)\), (d) \((100,200,200)\), and (e) \((200,200,100)\), respectively. Panels (f) and (g) show the results for \(n=3\) with \((L_1,L_2,L_3)=\) (f) \((100,200,200)\) and
(g) \((200,200,100)\), respectively. Panels (h)--(l) show the results for \(n=5\) with the same sets of
\((L_1,L_2,L_3)\) as panels (a)--(e), respectively. The blue, purple, black, green, and orange symbols (dashed curves) represent the simulation results (theoretical curves) of $\rho_{\rm br}(x)$ for $\rho_{\rm ini}\in\{0, 0.4, 0.5, 0.6, 1\}$.}
\label{fig:comparison_1n1_profiles_app}
\end{figure*}

Figure~\ref{fig:comparison_1n1_profiles_app} presents additional branch
density profiles of the \(1\)-\(n\)-\(1\) model for the same parameter
sets as Fig.~\ref{fig:comparison_1n1_phase_app}, complementing Fig.~\ref{fig:comparison_1n1_profiles}. The simulation results
are compared with the theoretical prediction for \(\rho_{\rm br}(x)\),
given by Eq.~\eqref{eq:rhobr_1n1}.  As in the main text, the profiles are nearly flat
in the LD and HD phases, whereas tilted profiles appear in the SW phase.

The simulation results reproduce the overall behavior predicted by the
theory for all parameter sets.  In particular, the tilted profiles vary
with \(\rho_{\rm ini}\) in the expected manner, confirming that the local
shock-wave picture applies also to these additional geometries.  Small
quantitative discrepancies remain in the spatially resolved profiles, as
discussed in the main text, but the leading behavior of
\(\rho_{\rm br}(x)\) is consistently captured by the coarse-grained
description.

\section{Estimation of \(\rho_-\) and \(\rho_+\) from Simulation Data}
\label{app:fitting_rhopm}

The numerical estimates of the boundary densities \(\rho_-\) and
\(\rho_+\) plotted as the simulation results in
Figs.~\ref{fig:comparison_rhopm_geometry} and
\ref{fig:comparison_rhopm_0n0} were obtained from \(\bar{\rho}_{\rm br}\). For each fixed geometry and each value
of \(n\), we first rounded the corresponding theoretical boundary density
\(\rho_{\rm c}^{\rm th}\) to three decimal places, where
\(\rho_{\rm c}^{\rm th}\) denotes either \(\rho_-\) or \(\rho_+\).  Around
this rounded value, denoted by \(\rho_{\rm c}^{(0)}\), we sampled eleven
initial densities,
\[
\rho_{\rm ini}=\rho_{\rm c}^{(0)}+0.01 k,
\qquad
k=-5,-4,\ldots,5 .
\]
The resulting data for \(\bar{\rho}_{\rm br}\) were then fitted as a
function of \(\rho_{\rm ini}\) by a piecewise-linear function.  This
procedure uses only the change in the mean branch density and does not
require a separate criterion for detecting a local shock.

For the lower boundary, only the data with
\(\rho_{\rm ini}\le 1/2\) were used.  We fitted them to
\[
f_-(\rho_{\rm ini};\rho_-,a_-)=
\begin{cases}
\rho_{\rm L}, & \rho_{\rm ini}<\rho_-,\\
\rho_{\rm L}+a_-(\rho_{\rm ini}-\rho_-), & \rho_{\rm ini}\ge \rho_-,
\end{cases}
\]
where \(\rho_{\rm L}\) is the theoretical branch density in the LD
plateau and \(a_-\) is a fitting parameter.  The best-fit value of
\(\rho_-\) was defined as the numerical lower boundary.

Similarly, for the upper boundary, only the data with
\(\rho_{\rm ini}\ge 1/2\) were used.  We fitted them to
\[
f_+(\rho_{\rm ini};\rho_+,a_+)=
\begin{cases}
\rho_{\rm H}+a_+(\rho_{\rm ini}-\rho_+), & \rho_{\rm ini}\le \rho_+,\\
\rho_{\rm H}, & \rho_{\rm ini}>\rho_+,
\end{cases}
\]
where \(\rho_{\rm H}\) is the theoretical branch density in the HD
plateau and \(a_+\) is a fitting parameter.  The best-fit value of
\(\rho_+\) was defined as the numerical upper boundary.

The plateau densities were fixed to the theoretical values.  For the
\(1\)-\(n\)-\(1\) model, we used
\[
\rho_{\rm L}=\frac{1}{2n},
\qquad
\rho_{\rm H}=1-\frac{1}{2n},
\]
whereas for the \(0\)-\(n\)-\(0\) model, we used
\[
\rho_{\rm L}=\frac{1}{n},
\qquad
\rho_{\rm H}=1-\frac{1}{n}.
\]
For each fit, the parameters \(\rho_\pm\) and \(a_\pm\) were determined
by minimizing the sum of squared deviations between the simulation data
and the corresponding piecewise-linear function.

\begin{figure*}[t]
\centering
\includegraphics[width=\textwidth]{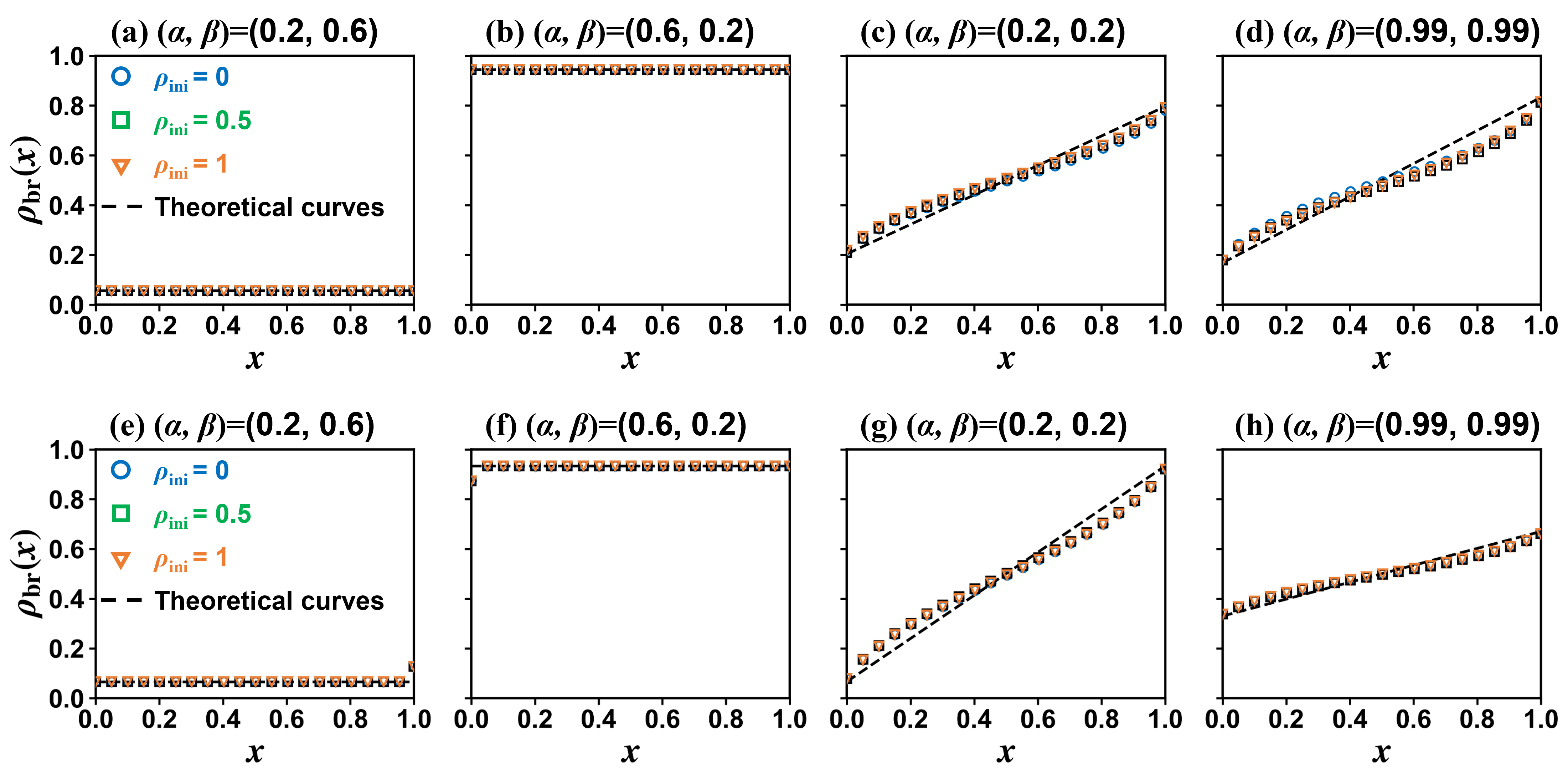}
\caption{Branch density profiles with $(\alpha,\beta)\ne(1,1)$.  
Panels (a)--(d) show the \(1\)-\(3\)-\(1\) model for $(\alpha, \beta)={\rm (a)\ } (0.2, 0.6), {\rm (b)\ } (0.6, 0.2), {\rm (c)\ } (0.2, 0.2), {\rm and \ (d)\ } (0.99, 0.99)$, respectively, with \((L_1,L_2,L_3)=(200,200,200)\). Panels (e)--(h) show the \(0\)-\(3\)-\(0\) model for the same values of $(\alpha, \beta)$ with \(L=200\). We note that for panels (c), (d), (g), and (h), the system is
first relaxed for $10^8$ time steps, and physical quantities are then obtained by averaging over $10^8$ time steps.
The blue, green, and orange symbols represent the simulation results of $\rho_{\rm ini}\in\{0, 0.5, 1\}$, respectively. The black dashed curves represent the theoretical estimates.}
\label{fig:alphabetaneq1}
\end{figure*}

\section{Branch Density Profiles with $(\alpha,\beta)\ne(1,1)$}
\label{sec:alphabetaneq1}

In the main text, we focus on the deterministic case
of $\alpha=\beta=1$, where the branch density can retain a memory of $\rho_{\rm ini}$.  In this appendix, we show that such
$\rho_{\rm ini}$-dependence does not persist for $(\alpha,\beta)\ne(1,1)$. 

\subsection{$1$-$n$-$1$ model}
\label{sec:alphabetaneq1_1n1}

For the $1$-$n$-$1$ model, the theoretical density profile can be obtained
by extending the domain-wall argument in Ref.~\cite{PhysRevE.108.044121}
to the case with $n$ parallel branches, and by taking the special case of
no friction at the merging point. Specifically, $\rho_{\rm br}(x)$ can be written as
\begin{widetext}
\begin{equation}
\rho_{\rm br}(x)
=
\begin{cases}
\dfrac{\alpha}{n(1+\alpha)},
& {\rm for} \  \alpha<\beta,\\
1-\dfrac{\beta}{n(1+\beta)},
& {\rm for} \  \alpha>\beta,\\
\dfrac{\alpha}{n(1+\alpha)}
+
\dfrac{n+(n-2)\alpha}{n(1+\alpha)}
\dfrac{1-\alpha+\{n+(n-2)\alpha\}x}
{n+2+(n-4)\alpha},
& {\rm for} \   \alpha=\beta.
\end{cases}
\label{eq:rhobr_abn1}
\end{equation}
\end{widetext}

Figures~\ref{fig:alphabetaneq1} (a)--(d) shows $\rho_{\rm br}(x)$ for the $1$-$3$-$1$ model for various sets of $(\alpha, \beta)$. The symbols (dashed curves) represent simulation
results (theoretical profiles given by Eqs. (\ref{eq:rhobr_abn1})) for $\rho_{\rm ini}\in\{0, 0.5, 1\}$. The four parameter sets consist of the LD case in panel (a), the HD case in
panel (b), and the two coexistence lines in panels (c) and (d).

The simulation profiles for different $\rho_{\rm ini}$ collapse onto the same theoretical curves, indicating no $\rho_{\rm ini}$-dependence for $(\alpha,\beta)\ne(1,1)$.

\subsection{$0$-$n$-$0$ model}
\label{sec:alphabetaneq1_0n0}

For the $0$-$n$-$0$ model, the theoretical density profile can be obtained from Ref.~\cite{de1999exact}, replacing the left and right boundary densities with the effective values
$\alpha/n$ and $\beta/n$, respectively.  
Specifically, $\rho_{\rm br}(x)$ can be written as
\begin{equation}
\rho_{\rm br}(x)
=
\begin{cases}
\dfrac{\alpha}{n}, & {\rm for} \ \alpha<\beta,\\
1-\dfrac{\beta}{n}, & {\rm for} \ \alpha>\beta,\\
\dfrac{\alpha}{n}
+
\left(1-\dfrac{2\alpha}{n}\right)x, & {\rm for} \ \alpha=\beta .
\end{cases}
\label{eq:rhobr_abn1_0n0}
\end{equation}

Figures~\ref{fig:alphabetaneq1} (e)--(h) shows $\rho_{\rm br}(x)$ for the $0$-$3$-$0$ model for various sets of $(\alpha, \beta)$. The symbols (dashed curves) represent simulation
results (theoretical profiles given by Eqs. (\ref{eq:rhobr_abn1_0n0})) for $\rho_{\rm ini}\in\{0, 0.5, 1\}$. The four parameter sets consist of the LD case in panel (e), the HD case in
panel (f), and the two coexistence lines in panels (g) and (h).

As in the $1$-$n$-$1$ model, the simulation profiles for different $\rho_{\rm ini}$ collapse onto the same theoretical curves, again indicating no $\rho_{\rm ini}$-dependence for $(\alpha,\beta)\ne(1,1)$. We note that small deviations near the branch boundaries are attributed to boundary-layer effects, which are not included in the coarse-grained profiles in Eq.~(\ref{eq:rhobr_abn1_0n0}).

\bibliographystyle{apsrev4-2}
\bibliography{references}

\end{document}